\documentstyle[11pt]{article}
\renewcommand{\thesection}{\Roman{section}.}

\renewcommand{\theequation}{\arabic{section}.\arabic{equation}}

\refstepcounter{equation}
\makeatletter                                                           

\def\ps@headings{                                                       
 \def\@oddhead{\footnotesize\rm\hfill\runninghead\hfill}
 \def\@evenhead{\@oddhead}                                              
 \def\@oddfoot{\rm\hfill\thepage\hfill}\def\@evenfoot{\@oddfoot} }

\makeatother                                                            

\begin{document}
\title{Wigner Function and Quantum Kinetic Theory
in Curved Space--Time and External Fields}
\def\runninghead{WIGNER FUNCTION AND QUANTUM KINETIC...}
\author{Oleg A. Fonarev
\\ {\em Racah Institute of Physics, The Hebrew University}
\\ {\em Jerusalem 91904, Israel}
\\ {\em E--mail: OLG@vms.huji.ac.il}}
\date{}
\pagestyle{headings}                                                   
\flushbottom                                                           
\maketitle
\begin{abstract}
A new definition of the Wigner function for quantum fields coupled to curved
space--time and an external Yang--Mills field is studied on the example of a
scalar and a Dirac fields. The definition uses the formalism of the tangent
bundles and is explicitly covariant and gauge invariant. Derivation of
collisionless quantum kinetic equations is carried out for both quantum fields
by using the first order formalism of Duffin and Kemmer. The evolution of the
Wigner function is governed by the quantum corrected Liouville--Vlasov equation
supplemented by the generalized mass--shell constraint. The structure of the
quantum corrections is perturbatively found in all adiabatic orders. The lowest
order quantum--curvature corrections coincide with the ones found by Winter.
\end{abstract}
\vspace{2\baselineskip}
PACS Indices: 05.30.--d; 05.20.Dd; 04.90.$+e$

\newpage

\section{Introduction}
\setcounter{equation}{0}
\hspace*{.5in}
The Wigner--function method is well--known as the most effective bridge between
quantum and macroscopic regimes. It allows one to derive from a quantum wave
equation a Liouville equation for a quantum distribution function and explore
the semiclassical limit, providing thus, a powerful ground for kinetic theory.
Recently, the Wigner transformation in curved space--time was studied in
different approaches$^{\ref{kn:winter2}-\ref{kn:graziani}}$.
In the present work, a covariant and gauge invariant Wigner function for
quantum fields coupled to external gravitational and Yang--Mills fields is
defined by using the horizontal lift of the derivative operator in the tangent
bundle. The tangent and the cotangent bundles arise naturally in general
relativistic kinetic theory and it was Vlasov, who first recognized
this$^{\ref{kn:vlasov}}$. This paper demonstrates, how the formalism of the
bundles allows one to derive quantum Liouville equations explicitly in all
orders of the perturbation theory. \\
\hspace*{.5in}
Sec.\ ~\ref{sec-defs} recalls some features of general relativistic kinetic
theory and gives the definition of a covariant Wigner function, proceeding from
considering expressions for dynamical observables (such as the current density
vector, the stress--energy tensor and so on) of quantum fields. In Sec.\
{}~\ref{sec-defs}~\ref{sec-duf} the first order formalism of Duffin and Kemmer
is generalized to curved space--time and it is shown that a scalar field in the
first order formalism is formally equivalent to
a Dirac field. This allows one to derive quantum Liouville equations for
different quantum fields in a universal fashion.
 Sec.\ ~\ref{sec-kin} derives quantum kinetic equations and generalized
mass--shell constraints for the Wigner functions in terms of a semiclassical
expansion. It is shown, by using the method of Ref.\ \ref{kn:hu}, how to
eliminate the mass--shell constraint to get a transport equation for a reduced
function on the classical mass--shell.
In Sec.\ ~\ref{sec-final} some unsolved problems are mentioned. The Appendices
contain some technical details and useful formulae.
\\ \hspace*{.5in}
Notations and conventions for a curvature tensor and a Yang--Mills field
throughout are such that the commutator of two covariant and gauge invariant
derivatives is
\begin{equation}
\left[ \tilde{\nabla}_{\alpha} ,
\tilde{\nabla}_{\beta} \right] \,
X_{\mu} = R_{\alpha\beta\mu\nu} \, X^{\nu} + \frac{ie}{\hbar} \,
F_{\alpha\beta} \, X_{\mu} \; ,
\end{equation}
where $e$ is a gauge charge and $\hbar$ is Planck's constant (the signature for
a metric is $(+---)$).
\section{Field equations and Wigner functions}
\setcounter{equation}{0}
\hspace*{.5in}
I \label{sec-defs} will consider both a scalar quantum field and a Dirac field
coupled to external gravitational and Yang--Mills fields. Let's consider first
a scalar field. The model is described by the quantum field equation (see, for
instance, Ref.\ \ref{kn:birdav}):
\begin{equation}
\left({g^{\alpha\beta}}(x) \tilde{\nabla}_{\alpha} \tilde{\nabla}_{\beta} +
m^{2}/\hbar^{2} -
\xi {R}(x)\right){{\bf \varphi}}(x)= V_{int}[{\bf \varphi}] \; ,
\label{eq:phi}
\end{equation}
the Einstein equations:
\begin{equation}
{R_{\alpha\beta}}(x) - \frac{1}{2} {R}(x) {g_{\alpha\beta}}(x) = k \langle{{\bf
T}_{\alpha\beta}}(x)\rangle + k {T_{\alpha\beta}^{YM}}(x) \; , \label{eq:ein}
\end{equation}
and the Yang--Mills equations:
\begin{equation}
\tilde{\nabla}_{\alpha} {F^{\alpha\beta}}(x) =  e \langle{{\bf
J}^{\beta}}(x)\rangle \; .          \label{eq:yan}
\end{equation}
Here
 $\xi$ is the nonminimal gravitational coupling constant,
${T_{\alpha\beta}^{YM}}(x)$ is the Yang--\-Mills field's stress--energy tensor
and $\langle{{\bf J}^{\alpha}}(x)\rangle$ and $\langle{{\bf
T}_{\alpha\beta}}(x)\rangle$ are the ensemble averaged current density vector
and stress--energy tensor of the scalar quantum field ${{\bf \varphi}}(x)$:
\begin{equation}
\langle {{\bf J}_{\alpha}}(x)\rangle = (\frac{i\hbar}{2}) \, \langle {{\bf
\varphi}^{\dagger}}(x) \stackrel{\leftrightarrow}{\tilde{\nabla}}_{\alpha}
{{\bf \varphi}}(x) \rangle \; , \label{eq:curq}
\end{equation}
\begin{eqnarray}
\langle {{\bf T}_{\alpha\beta}}(x)\rangle = (\frac{i\hbar}{2})^{2} \, \langle
{{\bf \varphi}^{\dagger}}(x)
\stackrel{\leftrightarrow}{\tilde{\nabla}}_{\alpha}
\stackrel{\leftrightarrow}{\tilde{\nabla}}_{\beta} {{\bf \varphi}}(x)\rangle +
\hspace*{.7in} \nonumber \\
 \mbox{} + \hbar^{2}\left(\xi{R_{\alpha\beta}}(x) + (\frac{1}{4} -
\xi)(\nabla_{\alpha}\nabla_{\beta} - {g_{\alpha\beta}}(x)\Box)\right) \langle
{{\bf \varphi}^{\dagger}}(x) {{\bf \varphi}}(x) \rangle +
 \langle {{\bf T}_{\alpha\beta}^{int}}(x) \rangle \; . \label{eq:enerq}
\end{eqnarray}
The last term in Eq.\ (\ref{eq:enerq}) depends on quantum interactions (the
right hand side of Eq.\ (\ref{eq:phi})), ${{\bf \varphi}^{\dagger}}(x)$ is a
conjugate field and
\begin{equation}
\stackrel{\leftrightarrow}{\tilde{\nabla}}_{\alpha} =
\stackrel{\rightarrow}{\tilde{\nabla}}_{\alpha} -
\stackrel{\leftarrow}{\tilde{\nabla}}_{\alpha} \; ,
\end{equation}
the arrows indicate the directions of acting of the derivative operator. \\
\hspace*{.5in}
 The ensemble averaging means that we have specified in-particles in an
asymptotic past (with not necessary vanishing curvature, but admitting a
reasonable definition of particles) and a density matrix ${\bf \rho}$ based on
in-states. Then the ensemble averaging of an operator ${{\bf A}}(x)$ is
\begin{equation}
\langle{{\bf A}}(x)\rangle = Sp({\bf \rho}{{\bf A}}(x)) / Sp \, {\bf \rho} \; .
\label{eq:averA}
\end{equation}
Generally speaking, expressions like (\ref{eq:averA}) diverge and
one has to renormalize it. It can be done in terms of a Wigner function by
using, for example, a procedure like the BPHZ-procedure in quantum field
theory$^{\ref{kn:fonarevthes}}$ or the dimensional
regularization$^{\ref{kn:habib1}}$. We will return to this question elsewhere.
In this paper all quantum operators throughout are considered to be somehow
regularized.  \\
\hspace*{.5in}
 In the classical limit a system of scalar point particles is described by a
distribution function ${F}(x,p)$ which is a probability density in the phase
space. The crucial feature of general relativistic kinetic theory is that in
general relativity the structure of the phase space differs from the one in
nongravity physics (see, for instance, Ref.\ \ref{kn:stewart}).  \\
First, the second argument of a distribution function is a (co)tangent vector
on a space--time manifold $\cal M$, which belongs to the (co)tangent space or
the fibre ${\cal T}_{x}^{\ast}(\cal M)$ over point $x$ (see, for instance,
Ref.\ \ref{kn:yano}). Coordinate transformations in the manifold,
\begin{equation}
x^{\alpha'} = {x^{\alpha'}}(x) \; ,   \label{eq:transx}
\end{equation}
induce transformations in the fibre:
\begin{equation}
p_{\alpha'} = \frac{\partial x^{\alpha}}{\partial x^{\alpha'}} \: p_{\alpha} \;
. \label{eq:transp}
\end{equation}
The set of cotangent spaces over all points of the manifold forms the cotangent
bundle over $\cal M$ with $\cal M$ being a base space:
\begin{equation}
{{\cal T}^{\ast}}({\cal M}) = \bigcup_{x \in {\cal M}} {\cal T}^{\ast}_{x}
(\cal M) \; ,
\end{equation}
and the distribution function ${F}(x,p)$ is a scalar function on the base space
and the cotangent bundle, that means that under the transformations
(\ref{eq:transx}),
(\ref{eq:transp}) it behaves as the following:
\begin{equation}
{F}(x,p) = {F'}\left({x'}(x),\frac{\partial x}{\partial x'}\cdot p\right) \; .
\end{equation}
An invariant volume element on the base space is known (see, for instance,
Ref.\ \ref{kn:kobayashi}),
\begin{equation}
dV^{(4)} = (-{g}(x))^{1/2}\: dx^{0}\wedge dx^{1}\wedge dx^{2}\wedge dx^{3} \;
.\label{eq:volx}
\end{equation}
Analogously, an invariant volume element on the cotangent bundle is
\begin{equation}
d\Pi^{(4)} = (-{g}(x))^{-1/2}\: dp_{0}\wedge dp_{1}\wedge dp_{2}\wedge dp_{3}
\; , \label{eq:volp}
\end{equation}
but the order of integration over the phase space in general relativity is
crucial: to overcome an ambiguity, one should first integrate over the momentum
space at fixed points $x$ and then over the space--time manifold. In order to
define procedures like coarse graining, which needs integration over finite
space--time regions of a distribution function at fixed {\em momentum}
variables $p$, one has to set a way of comparing the energy--momentum vectors
of distant particles.
 \\
Secondly, the energy--momentum vector of a classical point particle,
$p_{\alpha}$, is constrained by the mass--shell equation:
\begin{equation}
{g^{\alpha \beta}}(x)\: p_{\alpha}\: p_{\beta} - m^{2} = 0 \; , \label{eq:mass}
\end{equation}
where $m$ is the particle's mass (we suppose that all particles are identical).
More over, the energy $p_{0}$ of a particle must be positive. So, the particle
number in a phase space volume element is
\begin{equation}
dN = {F}(x,p) {\theta}(p_{0}) {\delta}(p^{2} - m^{2}) d\Pi^{(4)} dV^{(4)} \; .
\label{eq:pnum}
\end{equation}
One can include $\theta$-- and $\delta$--functions to the definition of the
volume element on the physical sector of the phase space. Then after
integration over $p_{0}$ we will recover the 7--dimensional phase space. But as
one will see below, for quantum particles the mass--shell constraint is more
complicated and not only expressed by ${\delta}$--function. It is thus more
natural to define the classical distribution function as a distributed
function,
\begin{equation}
{f_{cl}}(x,p) = {F}(x,p) {\theta}(p_{0}) {\delta}(p^{2} - m^{2}) \; ,
\label{eq:fcl}
\end{equation}
which has to be recovered from a quantum distribution function in the classical
limit.
If particles couple to external (selfconsistent) gravitational and
electromagnetic fields, the Liouville's equation which governs the dynamics of
the distribution function is then$^{\ref{kn:stewart}}$
\begin{equation}
( p^{\alpha} D_{\alpha} - e F_{\alpha\beta}\: p^{\alpha}\:
\frac{\partial}{\partial p_{\alpha}} )\: {f_{cl}}(x,p) = C[f_{cl}] \; ,
\label{eq:vla}
\end{equation}
where
\begin{equation}
D_{\alpha} = \nabla_{\alpha} + \Gamma_{\alpha\beta}^{\gamma}\: p_{\gamma}\:
\frac{\partial}{\partial p_{\beta}}  \; ,  \label{eq:dp}
\end{equation}
$\Gamma_{\alpha\beta}^{\gamma}$ are the Christoffel symbols and $C[f]$ is a
collision integral depended on the mechanism of particles interactions. The
operator $D_{\alpha}$ is called the horizontal lift of the derivative operator
in the cotangent bundle$^{\ref{kn:yano}}$ and it was introduced to kinetic
theory by Vlasov$^{\ref{kn:vlasov}}$. It is invariant under the transformations
(\ref{eq:transx}),(\ref{eq:transp}).
\\
The current density vector and the stress--energy tensor are expressed in terms
of moments of the distribution function in the momentum space:
\begin{equation}
\langle {J_{\alpha}}(x) \rangle = \int \frac{d^{4}p}{\sqrt{-{g}(x)}}\:
p_{\alpha}\: {f_{cl}}(x,p)  \; ,     \label{eq:cur}
\end{equation}
\begin{equation}
\langle {T_{\alpha\beta}}(x) \rangle = \int \frac{d^{4}p}{\sqrt{-{g}(x)}}\:
p_{\alpha} p_{\beta} {f_{cl}}(x,p) + \langle {T^{int}_{\alpha\beta}}(x) \rangle
 \; ,  \label{eq:ener}
\end{equation}
where the second term in Eq.\ (\ref{eq:ener}) depends on particles
interactions.
\\
\hspace*{.5in}
The goal of kinetic theory is to replace the field equation (\ref{eq:phi}) by a
kinetic equation for a quantum distribution function and express the
expectation values (\ref{eq:curq}),(\ref{eq:enerq}) in terms of this function,
so that in the classical limit (which has to be defined properly) we will
restore all the set of Eqs.\ (\ref{eq:fcl})--(\ref{eq:ener}). The most
convenient method for exploring the classical limit is the Wigner--function
one$^{\ref{kn:wigner}}$. In the special relativistic quantum field theory the
(oneparticle) Wigner function is the Fourier transform of a Green function with
respect to the difference of its coordinates$^{\ref{kn:degroot}}$:
\begin{equation}
{f}(x,p)=(\pi\hbar)^{-4}\:\int
d^{4}s\:e^{-2is^{\alpha}p_{\alpha}/\hbar}\langle{{\bf \varphi}}(x_{1}){{\bf
\varphi}^{\dagger}}(x_{2}) \rangle \; , \label{eq:wigm}
\end{equation}
where
\begin{equation}
s^{\alpha}=\frac{1}{2}(x^{\alpha}_{1}-x^{\alpha}_{2}) \; ,   \label{eq:dif}
\end{equation}
\begin{equation}
x^{\alpha}=\frac{1}{2}(x^{\alpha}_{1}+x^{\alpha}_{2})  \; .   \label{eq:mid}
\end{equation}
\hspace*{.5in} In general relativity the situation becomes ambiguous because
the definitions of the difference of coordinates (\ref{eq:dif}) and a middle
point (\ref{eq:mid}) are not invariant under transformations of coordinates and
there is no natural covariant generalization of the Fourier transformation.
The first attempt to generalize the definition (\ref{eq:wigm}) to general
relativity and to evaluate the lowest order curvature corrections to the Vlasov
equation was made by Winter$^{\ref{kn:winter2}}$.
His definition is based on considering a geodesic connecting two points $x_{1}$
and $x_{2}$. For evaluating the quantum corrections to the Vlasov equation   in
Winter's method one has to solve the geodesic and the geodesic deviation
equations, which is a very complicated problem even in the lowest adiabatic
order.
\\ \hspace*{.5in}
A different approach$^{\ref{kn:hu}}$ is based on introducing Riemann normal
coordinates centered at any point in the vicinity of $x_{1}$ and $x_{2}$, on
defining the difference of coordinates and the middle point by Eqs.\
(\ref{eq:dif}),(\ref{eq:mid}) in this coordinate system and on applying a
certain differential operator on the Wigner function, in order to remove the
dependence on the choice of a center of Riemann coordinates.
 This method is also covariant and is equivalent to the Winter's one, but it
saves much work while deriving the quantum corrections to the Vlasov equation.
\\ \hspace*{.5in}
An alternative approach was proposed in Ref.\ \ref{kn:fonarev1}, which uses the
formalism of the tangent bundles. This approach is explicitly covariant and
also gauge invariant from the beginning and allows one to evaluate the quantum
corrections to the kinetic equation up to any adiabatic order with few efforts.
I will recall the main idea of Ref.\ \ref{kn:fonarev1}. \\
\hspace*{.5in}
Consider first the expression for the current density vector (\ref{eq:curq}).
Let me rewrite this expression in the following way:
\begin{equation}
\langle {{\bf J}_{\alpha}}(x) \rangle = \frac{i\hbar}{2}\: \int_{{\cal
T}_{x}(\cal M)} d^{4}y \: \left(\frac{\partial}{\partial y^{\alpha}}\:
{\delta^{4}}(y)\right)\: \langle Tr\, {{\bf \Phi}}(x,-y) {{\bf
\Phi}^{\dagger}}(x,y) \rangle \; , \label{eq:curt}
\end{equation}
where $y^{\alpha}$ is a tangent vector belonging to the tangent space ${\cal
T}_{x}(\cal M)$ at a point $x$, ${\delta^{4}}(y)$ is the 4--dimensional
$\delta$--function, the trace is taken in the gauge group's representation
space and ${{\bf \Phi}}(x,-y)$ and ${{\bf \Phi}^{\dagger}}(x,y)$ are the
following series in the operators ${\tilde{\nabla}}_{\alpha}$ acting to the
fields ${{\bf \varphi}}(x)$ and ${{\bf \varphi}^{\dagger}}(x)$ {\em at a
point\/} $x$:
\begin{equation}
{{\bf \Phi}}(x,-y) = \left( 1 - y^{\alpha}{\tilde{\nabla}}_{\alpha} +
\frac{1}{2!\
}y^{\alpha}y^{\beta}{\tilde{\nabla}}_{\alpha}{\tilde{\nabla}}_{\beta} -
\frac{1}{3!\
}y^{\alpha}y^{\beta}y^{\gamma}{\tilde{\nabla}}_{\alpha}{\tilde{\nabla}}_{\beta}{\tilde{\nabla}}_{\gamma} + \ldots \right){{\bf \varphi}}(x)    \label{eq:ser1}
\end{equation}
\begin{equation}
{{\bf \Phi}^{\dagger}}(x,y) = \left( 1 + y^{\alpha}{\tilde{\nabla}}_{\alpha} +
\frac{1}{2!\
}y^{\alpha}y^{\beta}{\tilde{\nabla}}_{\alpha}{\tilde{\nabla}}_{\beta} +
\frac{1}{3!\
}y^{\alpha}y^{\beta}y^{\gamma}{\tilde{\nabla}}_{\alpha}{\tilde{\nabla}}_{\beta}{\tilde{\nabla}}_{\gamma} + \ldots \right){{\bf \varphi}^{\dagger}}(x)    \label{eq:ser2}
\end{equation}
The $\delta$--function in Eq.\ (\ref{eq:curt}) can be represented by the
standard way in a covariant manner, if one integrates over the cotangent space
${\cal T}^{\ast}_{x}(\cal M)$ at a point $x$:
\begin{equation}
{\delta^{4}}(y) = (\pi\hbar)^{-4}\int_{{\cal T}^{\ast}_{x}(\cal M)}
d^{4}p\:e^{-2iy^{\alpha}p_{\alpha}/\hbar}  \; .    \label{eq:del}
\end{equation}
Substituting the last expression into Eq.\ (\ref{eq:curt}) gives the
representation for the current density vector as an integral over the momentum
space:
\begin{equation}
\langle {{\bf J}_{\alpha}}(x) \rangle = \int\frac{d^{4}\,p}{\sqrt{-{g}(x)}}\:
p_{\alpha}\:Tr{f}(x,p) \; ,        \label{eq:curw}
\end{equation}
with the scalar function ${f}(x,p)$ being
\begin{equation}
{f}(x,p) = (\pi\hbar)^{-4}\: \sqrt{-{g}(x)}\:\int
d^{4}\,y\:e^{-2iy^{\alpha}p_{\alpha}/\hbar}\:\langle {{\bf \Phi}}(x,-y){{\bf
\Phi}^{\dagger}}(x,y)\rangle \; . \label{eq:wigf}
\end{equation}
The expression
(\ref{eq:curw})
for the current density vector
looks like the expression (\ref{eq:cur}) in the kinetic theory, this indicates
that we are on the right path. To be more sure, I will also represent the
stress--energy tensor (\ref{eq:enerq}) in terms of the function
(\ref{eq:wigf}). It can be done by the same way as for the current density
vector by noting that an item
\begin{equation}
(\frac{i\hbar}{2})^{n}\:\langle {{\bf \varphi}^{\dagger}}(x)\:
\stackrel{\leftrightarrow}{\tilde{\nabla}}_{(\alpha_{1}} \:
\stackrel{\leftrightarrow}{\tilde{\nabla}}_{\alpha_{2}}
\ldots
\stackrel{\leftrightarrow}{\tilde{\nabla}}_{\alpha_{n})}\:
{{\bf \varphi}}(x)\rangle \; ,    \label{eq:item}
\end{equation}
where the indices are symmetrized with $\frac{1}{n!\ }$--multiplier,
can be represented by the same steps (\ref{eq:curt})--(\ref{eq:del}) as the
following:
\begin{equation}
\int\frac{d^{4}\,p}{\sqrt{-{g}(x)}}\: p_{\alpha_{1}}\,p_{\alpha_{2}}\ldots
p_{\alpha_{n}}\:Tr{f}(x,p)  \; ,      \label{eq:itemw}
\end{equation}
with the same function (\ref{eq:wigf}). After that, we can write the
noninteracting part of the stress--energy tensor (\ref{eq:enerq}) as an
integral over the momentum space:
\begin{eqnarray}
\langle {{\bf T}_{\alpha\beta}}(x) \rangle =
\int\frac{d^{4}\,p}{\sqrt{-{g}(x)}}\: p_{\alpha}\,p_{\beta}\:Tr{f}(x,p) +
\hspace*{1in} \nonumber \\
 \mbox{} + \hbar^{2}\left(\xi{R_{\alpha\beta}}(x) + (\frac{1}{4} -
\xi)(\nabla_{\alpha}\nabla_{\beta} - {g_{\alpha\beta}}(x)\Box)\right)
\int\frac{d^{4}\,p}{\sqrt{-{g}(x)}}\: Tr{f}(x,p) +
 \langle {{\bf T}_{\alpha\beta}^{int}}(x) \rangle \; . \label{eq:enerw}
\end{eqnarray}
If one compares the expressions (\ref{eq:curw}) and (\ref{eq:enerw}) with
(\ref{eq:cur}),(\ref{eq:ener}), one can guess that in the classical limit the
function $Tr{f}(x,p)$ will tend to the classical distribution function
(\ref{eq:fcl}).
\\ \hspace*{.5in}
The classical limit for the nonrelativistic Wigner function is usually explored
from its transport equation (see, for instance, Ref.\ \ref{kn:carruthers}). So
now, the next step is to evaluate the general relativistic quantum kinetic
equation.
But the definition (\ref{eq:wigf}) for a general relativistic Wigner function
is still useless because of the expansions (\ref{eq:ser1}),(\ref{eq:ser2}).
The covariant derivative operator $\tilde{\nabla}_{\alpha}$ doesn't annihilate
the vector $y^{\beta}$ and one has to work a little harder to collect the
series in Eqs.\ (\ref{eq:ser1}),(\ref{eq:ser2}) into a convenient operator, as
it was in nongravity physics (compare with Ref.\ \ref{kn:degroot}). To do that,
let me introduce the horizontal lift of the covariant derivative on the {\em
tangent\/} bundle$^{\ref{kn:yano}}$:
\begin{equation}
\hat{\tilde{\nabla}}_{\alpha} = \tilde{\nabla}_{\alpha} -
\Gamma_{\alpha\gamma}^{\beta} \: y^{\gamma} \: \frac{\partial}{\partial
y^{\beta}}    \label{eq:dy}
\end{equation}
(compare with Eq.\ (\ref{eq:dp})). It can be easily shown that the operator
$\hat{\tilde{\nabla}}_{\alpha}$ does annihilate $y^{\beta}$ and so, we can
rewrite the n--th term of the series (\ref{eq:ser1}),(\ref{eq:ser2}) in the
following way:
\begin{equation}
\frac{1}{n!\ } \: y^{\alpha_{1}} \ldots y^{\alpha_{n}}
\tilde{\nabla}_{\alpha_{1}} \ldots \tilde{\nabla}_{\alpha_{n}} = \frac{1}{n! }
\: \left( y^{\alpha} \hat{\tilde{\nabla}}_{\alpha} \right)^{n} \; .
\end{equation}
After that we get compact expressions:
\begin{equation}
{{\bf \Phi}}(x,-y) = \exp(-y^{\alpha} \hat{\tilde{\nabla}}_{\alpha}) \, {{\bf
\varphi}}(x) \; ,       \label{eq:Phi1}
\end{equation}
\begin{equation}
{{\bf \Phi}^{\dagger}}(x,y) = \exp(y^{\alpha} \hat{\tilde{\nabla}}_{\alpha}) \,
{{\bf \varphi}^{\dagger}}(x) \; .       \label{eq:Phi2}
 \end{equation}
Eqs.\ (\ref{eq:wigf}),(\ref{eq:Phi1}) and (\ref{eq:Phi2}) give us the
definition of a covariant and gauge invariant Wigner function. As one will see
below, this definition is very convenient for evaluating the quantum kinetic
equation, but it looks rather formal if one wishes to compute the Wigner
function explicitly for a concrete system. It is shown in ~\ref{sec-A} that the
Wigner function (\ref{eq:wigf}) can be expressed in terms of base space
quantities only, and conjunctions with other approaches are discussed.
\\ \hspace*{.5in}
Let's now consider a Dirac field ${{\bf \psi}}(x)$ coupled to external
gravitational and Yang--Mills fields and satisfying the generalized Dirac
equation (see, for instance, Ref.\ \ref{kn:birdav}):
\begin{equation}
\left(i \hbar \, {\gamma^{\alpha}}(x) \, \tilde{\nabla}_{\alpha} - m \right) \,
{{\bf \psi}}(x) = V_{int}[{\bf \psi}] \; ,   \label{eq:Dir1}
\end{equation}
where ${\gamma^{\alpha}}(x)$ are the Dirac matrices in curved space--time
obeying the (anti)commutative rule:
\begin{equation}
{\gamma^{\alpha}}(x) {\gamma^{\beta}}(x) + {\gamma^{\beta}}(x)
{\gamma^{\alpha}}(x) = 2 {g^{\alpha\beta}}(x) \; , \label{eq:comd}
\end{equation}
and the acting of the derivative operator to a Dirac field is determined by the
Fock--Ivanenko connection$^{\ref{kn:fock}} {\Omega_{\alpha}}(x)$ :
\begin{equation}
\tilde{\nabla}_{\alpha}\, {{\bf \psi}}(x) = \frac{\partial}{\partial
x^{\alpha}}\, {{\bf \psi}}(x) - {\Omega_{\alpha}}(x) {{\bf \psi}}(x) +
\frac{ie}{\hbar} {A_{\alpha}}(x) {{\bf \psi}}(x)    \label{eq:nab1}
\end{equation}
(${A_{\alpha}}(x)$ is the Yang--Mills potential).
\\ A Dirac conjugate field
${\bar{\bf \psi}}(x)={{\bf \psi}^{\dagger}}(x) {\gamma}(x)$, with ${\gamma}(x)$
commuting with all ${\gamma^{\dagger\alpha}}(x)$, satisfies the equation:
\begin{equation}
{\bar{\bf \psi}}(x) \, \left(- i \hbar \, {\gamma^{\alpha}}(x) \,
\stackrel{\leftarrow}{\tilde{\nabla}}_{\alpha} - m \right) =
\overline{V_{int}[{\bf \psi}]} \; ,   \label{eq:Dir2}
\end{equation}
with
\begin{equation}
\tilde{\nabla}_{\alpha}\, {\bar{\bf \psi}}(x) = \frac{\partial}{\partial
x^{\alpha}}\, {\bar{\bf \psi}}(x) + {\bar{\bf \psi}}(x){\Omega_{\alpha}}(x) -
\frac{ie}{\hbar} {\bar{\bf \psi}}(x) {A_{\alpha}}(x) \; .    \label{eq:nab2}
\end{equation}
\\ \hspace*{.5in}
A covariant and gauge invariant Wigner function for a Dirac field can be
defined by the same trick as for the scalar field. It is the following (matrix)
function (the sign minus appears here because of the Fermi--statistics):
\begin{equation}
{N}(x,p) = -(\pi\hbar)^{-4}\: \sqrt{-{g}(x)}\:\int
d^{4}\,y\:e^{-2iy^{\alpha}p_{\alpha}/\hbar}\:\langle {{\bf
\Psi}}(x,-y){\bar{\bf \Psi}}(x,y)\rangle \; , \label{eq:wigN}
\end{equation}
where the functions ${{\bf \Psi}}(x,-y)$ and ${\bar{\bf \Psi}}(x,y)$ are
related to the Dirac field ${{\bf \psi}}(x)$ and the conjugate field ${\bar{\bf
\psi}}(x)$ in the same way as for a scalar field (Eqs.\
(\ref{eq:Phi1}),(\ref{eq:Phi2}) and (\ref{eq:dy}) ):
\begin{equation}
{{\bf \Psi}}(x,-y) = \exp(-y^{\alpha} \hat{\tilde{\nabla}}_{\alpha}) \, {{\bf
\psi}}(x) \; ,       \label{eq:Psi1}
\end{equation}
\begin{equation}
{\bar{\bf \Psi}}(x,y) = \exp(y^{\alpha} \hat{\tilde{\nabla}}_{\alpha}) \,
{\bar{\bf \psi}}(x)  \; .      \label{eq:Psi2}
\end{equation}
\hspace*{.5in} In Ref.\ \ref{kn:fonarev1} the noninteracting part of the
stress--energy tensor, the current density vector and the spin tensor of a
Dirac field are expressed by using the generalized Gordon decomposition (see,
for instance, Ref.\ \ref{kn:bjorken}), in terms of the independent components
of the Wigner function (\ref{eq:wigN}):
\begin{equation}
{A}(x,p) = m^{-1} \, Tr \, {N}(x,p) \; ,      \label{eq:wigsc}
\end{equation}
\begin{equation}
{A^{\alpha\beta}}(x,p) = \frac{1}{2} i m^{-1} \, Tr \left( {\gamma^{[
\alpha}}(x) {\gamma^{\beta ]}}(x) {N}(x,p) \right)  \; ,     \label{eq:wigte}
\end{equation}
\begin{equation}
{B}(x,p) = -i m^{-1} \, Tr \left( \gamma^{5} {N}(x,p) \right) \; ,
\label{eq:wigpsc}
\end{equation}
where $\gamma^{5} = -i/4!\ \sqrt{-{g}(x)} \varepsilon_{\alpha\beta\mu\nu}
{\gamma^{\alpha}}(x) {\gamma^{\beta}}(x) {\gamma^{\mu}}(x) {\gamma^{\nu}}(x)$.
\\ \hspace*{.5in}
At the end of this section, I would like to note that a coupling to an external
scalar field ${\phi}(x)$ can be described formally by changing the mass terms
in Eqs.\ (\ref{eq:phi}) and (\ref{eq:Dir1}),(\ref{eq:Dir2}):
\begin{equation}
m \longmapsto m + {\phi}(x) \; .
\end{equation}
Such coupling doesn't influence the definition of the Wigner functions but does
influence the kinetic equation.
\subsection{The covariant Duffin--Kemmer formalism}
\hspace*{.5in}
To \label{sec-duf} derive the quantum kinetic equation it will be convenient
for us to describe the scalar field within the first order formalism, in which
the field equation looks formally like the Dirac equation. In special
relativity this type of formalism was proposed by Duffin$^{\ref{kn:duffin}}$
and developed by Kemmer$^{\ref{kn:kemmer}}$. I will extend it to general
relativity. Let me introduce the new fields:
\begin{equation}
{{\bf u}_{\alpha}}(x) = i \hbar \mu^{-1/2} \, \tilde{\nabla}_{\alpha} {{\bf
\varphi}}(x)  \makebox[.5in]{;} {{\bf u}_{4}}(x) = \mu^{1/2} \, {{\bf
\varphi}}(x)  \; , \label{eq:newphi}
\end{equation}
where $\mu$ is an arbitrary parameter of the mass dimension. Uniting the fields
(\ref{eq:newphi}) into one five--component field:
\begin{equation}
{{\bf u}}(x) = \left( \begin{array}{c} {{\bf u}_{\alpha}}(x) \\ {{\bf
u}_{4}}(x) \end{array} \right)  \; ,   \label{eq:u}
\end{equation}
and defining the $5 \times 5$ --matrices:
\begin{equation}
{\Gamma^{\alpha}}(x) = \left( \begin{array}{ccc} O & | &  \delta^{\alpha}_{\nu}
\\ --- & | & - \\ {g^{\alpha\beta}}(x) & | & 0 \end{array}  \right)
\label{eq:gam} \; ,
\end{equation}
\begin{equation}
{M}(x) = \left( \begin{array}{ccc} \mu \, \delta^{\beta}_{\nu} & | & O \\ --- &
| & - \\ O & | & {m}(x) \end{array} \right) \; ,  \label{eq:Mas}
\end{equation}
with
\begin{equation}
{m}(x) = \mu^{-1} (m^{2} - \xi \hbar^{2} {R}(x))  \; ,  \label{eq:mas}
\end{equation}
one can rewrite the equation (\ref{eq:phi}) in a form which is formally the
same as for a Dirac field (\ref{eq:Dir1}) (with a matrix variable mass):
\begin{equation}
\left(i \hbar \, {\Gamma^{\alpha}}(x) \, \tilde{\nabla}_{\alpha} - {M}(x)
\right) \, {{\bf u}}(x) = V'_{int}[{\bf u}] \; ,   \label{eq:Duf1}
\end{equation}
the only difference is that the matrices ${\Gamma^{\alpha}}(x)$ obey the
Duffin--Kemmer commutative rule rather than the Dirac one (\ref{eq:comd}):
\begin{equation}
{\Gamma^{\alpha}}(x) {\Gamma^{\nu}}(x) {\Gamma^{\beta}}(x) +
{\Gamma^{\beta}}(x) {\Gamma^{\nu}}(x) {\Gamma^{\alpha}}(x) = {g^{\nu\alpha}}(x)
{\Gamma^{\beta}}(x) + {g^{\nu\beta}}(x) {\Gamma^{\alpha}}(x)  \; .
\label{eq:comG}
\end{equation}
The covariant differentiation in Eq.\ (\ref{eq:Duf1}) is defined by the
connection ${\Lambda_{\alpha}}(x)$:
\begin{equation}
{\Lambda_{\alpha}}(x) =
\left( \begin{array}{ccc} {\Gamma^{\beta}_{\alpha\nu}}(x) & | & O \\ --- & | &
- \\ O & | & 0 \end{array} \right) \; ,  \label{eq:Lambda}
\end{equation}
so that (compare with Eq.\ (\ref{eq:nab1}))
\begin{equation}
\tilde{\nabla}_{\alpha}\, {{\bf u}}(x) = \frac{\partial}{\partial x^{\alpha}}\,
{{\bf u}}(x) - {\Lambda_{\alpha}}(x) {{\bf u}}(x) + \frac{ie}{\hbar}
{A_{\alpha}}(x) {{\bf u}}(x) \; .   \label{eq:nabu1}
\end{equation}
If we define the matrix ${\Gamma}(x)$, which commutes with all
${\Gamma^{\dagger\alpha}}(x)$:
\begin{equation}
{\Gamma}(x) = \left( \begin{array}{ccc} {g^{\alpha\beta}}(x) & | & O \\ --- & |
& - \\ O & | & 1 \end{array} \right) \; ,   \label{eq:ga}
\end{equation}
then the conjugate field
\begin{equation}
{\bar{{\bf u}}}(x) = {{\bf u}^{\dagger}}(x) {\Gamma}(x) = \left( {{\bf
u}^{\dagger\alpha}}(x) , {{\bf u}^{\dagger}_{4}}(x) \right)
\end{equation}
satisfies the following equation:
\begin{equation}
{\bar{\bf u}}(x) \, \left(- i \hbar \, {\Gamma^{\alpha}}(x) \,
\stackrel{\leftarrow}{\tilde{\nabla}}_{\alpha} - {M}(x) \right) =
\overline{V'_{int}[{\bf u}]} \; .   \label{eq:Duf2}
\end{equation}
It can be checked that the covariant derivative operator annihilates the
$\Gamma$--matrices (\ref{eq:gam}) and (\ref{eq:ga}) as well as the Dirac
$\gamma$--matrices:
\begin{equation}
\tilde{\nabla}_{\alpha}\, {\Gamma^{\beta}}(x) = \frac{\partial}{\partial
x^{\alpha}}\, {\Gamma^{\beta}}(x) - \left[{\Lambda_{\alpha}}(x) ,
{\Gamma^{\beta}}(x) \right] + {\Gamma^{\beta}_{\alpha\nu}}(x) \,
{\Gamma^{\nu}}(x) = 0 \; , \label{eq:propG}
\end{equation}
\begin{equation}
\tilde{\nabla}_{\alpha}\, {\Gamma}(x) = \frac{\partial}{\partial x^{\alpha}}\,
{\Gamma}(x) - {\Lambda^{\dagger}_{\alpha}}(x) \, {\Gamma}(x) - {\Gamma}(x) \,
{\Lambda^{\alpha}}(x) = 0 \; .
\end{equation}
The analogy between the scalar field in the first order formalism and the Dirac
field will show itself more explicitly if one looks at the expressions for the
commutator of two covariant derivatives acting to each field. If one denotes:
\begin{equation}
{{\bf u}_{s}}(x) = \left\{ \begin{array}{ll}
{{\bf u}}(x)  & \mbox{for $s=1$  (the scalar field)} \\
{{\bf \Psi}}(x) & \mbox{for $s=1/2$  (the Dirac field)}  \end{array}  \right.
 \label{eq:us}
\end{equation}
and
\begin{equation}
{\Gamma_{s}^{\alpha}}(x) = \left\{ \begin{array}{ll}
{\Gamma^{\alpha}}(x)
 & \mbox{for $s=1$ } \\
{\gamma^{\alpha}}(x)
 & \mbox{for $s=1/2$ }  \end{array}  \right. \; ,   \label{eq:Gs}
\end{equation}
then it can be obtained easily from Eqs.\ (\ref{eq:nab1}),(\ref{eq:nabu1}) that
\begin{equation}
\left[ \tilde{\nabla}_{\alpha} ,
\tilde{\nabla}_{\beta} \right] \, {{\bf u}_{s}}(x) = {A_{s \alpha\beta}}(x) \,
{{\bf u}_{s}}(x) \; ,  \label{eq:comu1}
\end{equation}
with
\begin{equation}
{A_{s \alpha\beta}}(x) = s^{2} \, {R_{\alpha\beta\mu\nu}}(x) \,
{\Gamma_{s}^{\mu}}(x) \, {\Gamma_{s}^{\nu}}(x) + \frac{ie}{\hbar} \,
{F_{\alpha\beta}}(x) \; .       \label{eq:A}
\end{equation}
For the conjugate fields one gets:
\begin{equation}
\left[ \tilde{\nabla}_{\alpha} ,
\tilde{\nabla}_{\beta} \right] \, {\bar{\bf u}_{s}}(x) = - {\bar{\bf u}_{s}}(x)
\, {A_{s \alpha\beta}}(x) \; .   \label{eq:comu2}
\end{equation}
Let me now define the unified Wigner function:
\begin{equation}
{N_{s}}(x,p) = \epsilon_{s} \, (\pi\hbar)^{-4}\: \sqrt{-{g}(x)}\:\int
d^{4}\,y\:e^{-2iy^{\alpha}p_{\alpha}/\hbar}\:\langle {{\bf
U}_{s}}(x,-y){\bar{\bf U}_{s}}(x,y)\rangle \; , \label{eq:wigNs}
\end{equation}
with ${{\bf U}_{s}}(x,-y)$  and ${\bar{\bf U}_{s}}(x,y)$ being related to
${{\bf u}_{s}}(x)$ and ${\bar{\bf u}_{s}}(x)$ respectively by means of formulae
analogous to (\ref{eq:Phi1}),(\ref{eq:Phi2}) and
\begin{equation}
\epsilon_{s} = \left\{ \begin{array}{ll}
 1 & \mbox{for $s=1$ } \\
 -1 & \mbox{for $s=1/2$ }  \end{array}  \right. \; .   \label{eq:eps}
\end{equation}
Then for $s=1/2$ one will get the Wigner function of the Dirac field
(\ref{eq:wigN}) immediately. In order to extract the Wigner function of the
scalar field (\ref{eq:wigf}) from (\ref{eq:wigNs}), one should introduce the
projection operators:
\begin{equation}
{\cal P}_{1} = \frac{1}{3} \, \left( {\Gamma^{\alpha}}(x) \,
{\Gamma_{\alpha}}(x) - 1 \right) =
 \left( \begin{array}{ccc} O & | & O \\ --- & | & - \\ O & | & 1 \end{array}
\right)  \; ,  \label{eq:P1}
\end{equation}
\begin{equation}
{\cal P}_{2} = \frac{1}{3} \, \left(4 - {\Gamma^{\alpha}}(x) \,
{\Gamma_{\alpha}}(x) \right) =
 \left( \begin{array}{ccc} \delta^{\beta}_{\nu} & | & 0 \\ --- & | & - \\ 0 & |
& 0 \end{array} \right)  \; ,  \label{eq:P2}
\end{equation}
with the following properties:
\begin{equation}
{\cal P}_{1} + {\cal P}_{2} = 1 \makebox[.2in]{;} {\cal P}_{1} \: {\cal P}_{2}
= 0
\makebox[.2in]{;} {\cal P}_{i}^{2} = {\cal P}_{i} \; ,  \label{eq:prop1}
\end{equation}
\begin{equation}
{\cal P}_{1} \: {\Gamma^{\alpha}}(x) = {\Gamma^{\alpha}}(x) \: {\cal P}_{2}
\makebox[.2in]{;}
{\cal P}_{2} \: {\Gamma^{\alpha}}(x) = {\Gamma^{\alpha}}(x) \: {\cal P}_{1} \;
, \label{eq:prop2}
\end{equation}
\begin{equation}
 {\Gamma^{\alpha}}(x) \: {\Gamma^{\beta}}(x) \: {\cal P}_{1} =
{g^{\alpha\beta}}(x) \: {\cal P}_{1} \; .   \label{eq:prop3}
\end{equation}
Then
\begin{equation}
{f}(x,p) = \mu^{-1} \: tr \, \left( {\cal P}_{1} \: {N_{1}}(x,p) \right) \; .
\label{eq:extrf}
\end{equation}
\hspace*{.5in} In conclusion, it is worth noting that the Duffin--Kemmer
formalism can be introduced for a vector field also. The evaluation of a
quantum kinetic equation in this case was carried out in Ref.\
\ref{kn:fonarevthes}.
\section{Quantum kinetic equations and }
\section*{\hspace*{.2in}mass--shell constraints}
\setcounter{equation}{0}
\label{sec-kin}
\subsection{The first order equation for the unified Wigner function}
\hspace*{.5in} If one takes into account the formal analogy between the Dirac
field and the scalar field in the first order formalism, one can guess that the
equations which govern the dynamics of the Wigner functions, look formally
similar for both fields. These equations can be extracted from the equation for
the unified Wigner function (\ref{eq:wigNs}). Before proceeding to the
derivation of this equation, I will prove the following identity (see also
Ref.\ \ref{kn:ignat}):
\begin{equation}
\tilde{D}_{\alpha} \, {N_{s}}(x,p) = \epsilon_{s} \, (\pi\hbar)^{-4}\,
\sqrt{-{g}(x)}\,\int d^{4}\, y\, e^{-2iy^{\sigma}p_{\sigma}/\hbar}\,
\hat{\tilde{\nabla}}_{\alpha} \langle {{\bf U}_{s}}(x,-y){\bar{\bf
U}_{s}}(x,y)\rangle \; , \label{eq:iden1}
\end{equation}
where $\tilde{D}_{\alpha}$ is the horizontal lift of the derivative operator in
the {\em cotangent\/} bundle ( Eq.\ (\ref{eq:dp}) with the minimal gauge
invariant extension ) and $\hat{\tilde{\nabla}}_{\alpha}$ is the horizontal
lift of the derivative operator in the {\em tangent\/} bundle (\ref{eq:dy}). \\
The identity (\ref{eq:iden1}) can be easily proved if one works in the Riemann
normal coordinates centered at point $x ^{\ref{kn:kobayashi}}$. In this
coordinate system both the operator
$\tilde{D}_{\alpha}$ and the operator $\hat{\tilde{\nabla}}_{\alpha}$
coincide with the gauge invariant extension of the partial derivative operator
$\tilde{\partial}_{\alpha}$ and the identity (\ref{eq:iden1}) is trivial. The
proof will be completed if one takes into account the invariance of the
derivative operators
$\tilde{D}_{\alpha}$ and $\hat{\tilde{\nabla}}_{\alpha}$
and of the volume element in Eq.\ (\ref{eq:iden1}).\\
Utilizing now the identities (\ref{eq:Halpha}),(\ref{eq:dyU}) from
{}~\ref{sec-B}, we will get, after integrating once by parts:
\begin{eqnarray}
\tilde{D}_{\alpha} \, {N_{s}}(x,p) = \frac{2i}{\hbar} \, p_{\alpha} \,
{N_{s}}(x,p) +  \hspace*{1.5in} \nonumber \\
\mbox{} + \epsilon_{s} \, (\pi\hbar)^{-4}\, \sqrt{-{g}(x)}\,\int d^{4}\, y\,
e^{-2iy^{\sigma}p_{\sigma}/\hbar}\, \langle 2\left( e^{-y^{\nu}
\hat{\tilde{\nabla}}_{\nu}} \, \hat{\tilde{\nabla}}_{\alpha} \, {{\bf
u}_{s}}(x) \right) {\bar{\bf U}_{s}}(x,y) -
 {G_{s\alpha}}(x,y) \rangle  \; ,  \label{eq:iden2}
\end{eqnarray}
where
\begin{eqnarray}
{G_{s\alpha}}(x,y) = {{\bf U}_{s}}(x,-y) \left(
{\hat{G}_{\alpha}}(x,y){\bar{{\bf U}}_{s}}(x,y) \right) - \hspace*{.5in}
\nonumber \\
\mbox{} - \left( {\hat{G}_{\alpha}}(x,-y){{\bf U}_{s}}(x,-y) \right) {\bar{{\bf
U}}_{s}}(x,y) +
2 \left( {\hat{H}_{\alpha}}(x,-y){{\bf U}_{s}}(x,-y) \right) {\bar{{\bf
U}}_{s}}(x,y) \; .    \label{eq:G}
\end{eqnarray}
\hspace*{.5in} So far, the equality (\ref{eq:iden2}) is an identity. In order
to obtain an equation, let me multiply it by $i\hbar/2 \,
{\Gamma^{\alpha}_{s}}(x)$ ( with a summation over $\alpha$ ). Taking into
account the property (\ref{eq:propG}) and using the field equations
(\ref{eq:Dir1}) or (\ref{eq:Duf1}) together with the identity (\ref{eq:comM}),
one will get:
\begin{eqnarray}
{\hat{M}}(x,-\frac{i\hbar}{2}\frac{\partial}{\partial p}) \, {N_{s}}(x,p) =
{\Gamma_{s}^{\alpha}}(x) \left( p_{\alpha} + \frac{i\hbar}{2}
\tilde{D}_{\alpha} \right) {N_{s}}(x,p) + \hspace*{.2in} \nonumber \\
\mbox{} + \frac{i\hbar}{2} {\Gamma_{s}^{\alpha}}(x) \, \epsilon_{s} \,
(\pi\hbar)^{-4}\, \sqrt{-{g}(x)}\,\int d^{4}\, y\,
e^{-2iy^{\sigma}p_{\sigma}/\hbar}\, \langle {G_{s\alpha}}(x,y) \rangle  +
C_{s}^{col} \; , \label{eq:kin1}
\end{eqnarray}
where ${\hat{M}}(x,-\frac{i\hbar}{2}\frac{\partial}{\partial p})$ is given by
Eq.\ (\ref{eq:Mxy}) with the substitution:
\begin{equation}
y^{\alpha} \longmapsto \frac{i\hbar}{2}\frac{\partial}{\partial p_{\alpha}} \;
, \label{eq:subst}
\end{equation}
and $C_{s}^{col}$ is a "collision" term which depends on the interaction terms
in Eqs.\ (\ref{eq:Dir1}) and (\ref{eq:Duf1}). I will not discuss this term in
this paper but return to this question elsewhere. In the rest of the paper I
will omit the "collision" term and deal only with collisionless kinetic
equations.
\\  The equation (\ref{eq:kin1}) will be a kinetic equation
if one also expresses the second term on the right--hand side in terms of the
function ${N_{s}}(x,p)$. It can be done by using the results of ~\ref{sec-C}.
Substituting Eq.\ (\ref{eq:Gsery})
into the integrand in Eq.\ (\ref{eq:kin1}), using the identity (\ref{eq:iden1})
once again and integrating by parts, one will get the following (collisionless)
equation:
\begin{eqnarray}
{\hat{M}}(x,-\frac{i\hbar}{2}\frac{\partial}{\partial p}) \, {N_{s}}(x,p) =
{\Gamma_{s}^{\alpha}}(x) \left( \frac{i\hbar}{2} \hat{R}_{1\alpha}^{\nu}
\tilde{D}_{\nu} +\hat{R}_{2\alpha}^{\nu} \, p_{\nu} \right) {N_{s}}(x,p)  +
\nonumber \\
+ \frac{i\hbar}{2} {\Gamma_{s}^{\alpha}}(x) \hat{A}_{s1\alpha}\, {N_{s}}(x,p)
+ \frac{i\hbar}{2} {\Gamma_{s}^{\alpha}}(x) \, {N_{s}}(x,p) \,
\stackrel{\leftarrow}{\hat{A}}_{s2\alpha} \; . \hspace*{.2in}
\label{eq:kin2}
\end{eqnarray}
Here $\hat{R}_{1,2\alpha}^{\nu}$ and $\hat{A}_{s1,2\alpha}$ are represented by
the infinite series in the momentum derivatives:
\begin{equation}
\hat{R}_{1,2\alpha}^{\nu} = \delta_{\alpha}^{\nu} + \sum_{k=2}^{\cal 1}
\hat{K}_{k\alpha}^{1,2\beta} \hat{R}_{\beta}^{(k)\nu} \; , \label{eq:R12}
\end{equation}
\begin{equation}
\hat{A}_{s1,2\alpha} = \sum_{k=2}^{\cal 1} \hat{K}_{k\alpha}^{3,4\beta}
\hat{A}_{s\beta}^{(k-1)} \; , \label{eq:A12}
\end{equation}
with
\begin{equation}
\hat{K}_{k\alpha}^{i\beta} = \delta_{\alpha}^{\beta} C_{k}^{i} +
\sum_{n=1}^{\cal 1}\sum_{k_{1}=2}^{\cal 1}\ldots\sum_{k_{n}=2}^{\cal 1}
C_{kk_{1}\ldots k_{n}}^{i} \hat{R}_{\alpha}^{(k_{n})\beta_{n}}\ldots
\hat{R}_{\beta_{2}}^{(k_{1})\beta}  \label{eq:Ki}
\end{equation}
and $C_{k_{1}\ldots k_{n}}^{i}$ being numerical coefficients (their explicit
expressions are presented in ~\ref{sec-C}).
The operators $\hat{R}_{\alpha}^{(k)\beta}$ and $\hat{A}_{s\alpha}^{(k)}$ in
Eqs.\ (\ref{eq:R12})--(\ref{eq:Ki}) are defined as follows:
\begin{equation}
\hat{R}_{\alpha}^{(k)\beta} = (\frac{i\hbar}{2} )^{k} \,
R^{\beta}_{\nu_{1}\alpha\nu_{2};\nu_{3}\ldots\nu_{k}} \, \partial^{k}/\partial
p_{\nu_{1}}\ldots p_{\nu_{k}} \; ,     \label{eq:Rk}
\end{equation}
\begin{equation}
\hat{A}_{s\alpha}^{(k)} = (\frac{i\hbar}{2} )^{k} \,
A_{s\alpha\nu_{1};\nu_{2}\ldots\nu_{k}} \, \partial^{k}/\partial
p_{\nu_{1}}\ldots p_{\nu_{k}} \; ,     \label{eq:Ak}
\end{equation}
with $A_{s\alpha\nu}$ being defined by (\ref{eq:A}).
\\  \hspace*{.5in}
The equation (\ref{eq:kin2}) together with the definitions
(\ref{eq:R12})--(\ref{eq:Ak}) give us the quantum kinetic equation for the
unified Wigner function (\ref{eq:wigNs}) in the collisionless approximation.
 But this equation is not yet in a useful form for exploring quantum kinetic
properties. Firstly, the infinite series in Eqs.\ (\ref{eq:R12})--(\ref{eq:Ki})
are formal as long as nothing is known about their convergence.
The situation here is quite similar to the one arisen in the nonrelativistic
case (see, for instance, Ref.\ \ref{kn:carruthers}).
 One has to assume that quantum interference is suppressed somehow and that the
external fields are "slowly varying" at the effective Compton scale (compare
also with the discussion in Ref.\ \ref{kn:hu}).
 Then Eqs.\ (\ref{eq:R12})--(\ref{eq:Ki}) can be approximated by the first few
terms and Eq.\ (\ref{eq:kin2}) will be a differential equation of a finite
order.
On the other hand, in some cases the infinite series in the momentum
derivatives can be collected into integrals and the kinetic equation will be an
integro--differential equation$^{\ref{kn:carruthers}}$.
If the quantum system is in a highly coherent state then some selective
summation of infinite subseries could provide a reasonable
approximation$^{\ref{kn:heller}}$.
The problem is a serious one but its discussion will be reserved for further
study (for the recent discussion see Ref.\ \ref{kn:habib2}).
It is worth noting that similar expansions arise in the description of the
extended bodies in general relativity$^{\ref{kn:dixon}}$.
\\  Secondly, the equation (\ref{eq:kin2}) doesn't look like a transport
equation. It will do so if we  "square" it (compare with the evaluation of the
transport equation for 1/2--spin particles in Ref.\ \ref{kn:degroot}). For this
purpose let me write Eq.\ (\ref{eq:kin2}) in compact notations:
\begin{equation}
\hat{M} {N_{s}}(x,p) = \hat{K_{s}} {N_{s}}(x,p) \; ,    \label{eq:com}
\end{equation}
where by $\hat{K_{s}}$ I have denoted the operator acting on the function
${N_{s}}(x,p)$ on the right--hand side of Eq.\ (\ref{eq:kin2}).
\\  Later on, the scalar field and the Dirac field will be treated separately.
\subsection{The scalar field case}
\hspace*{.5in} An equation for the Wigner function of the scalar field
(\ref{eq:wigf}) can be obtained by the means of the prescription
(\ref{eq:extrf}). Multiplying the equation (\ref{eq:com}) from the left by the
projection operators (\ref{eq:P1}) and (\ref{eq:P2}), one gets, by using the
properties (\ref{eq:prop2}) and the explicit matrix form of Eq.\
(\ref{eq:kin2}), the following two equations:
\begin{equation}
{\cal P}_{1} \, \hat{M} {N_{1}}(x,p) = \hat{K_{1}} \, {\cal P}_{2} \,
{N_{1}}(x,p) \; ,    \label{eq:PM1}
\end{equation}
\begin{equation}
{\cal P}_{2} \, \hat{M} {N_{1}}(x,p) = \hat{K_{1}} \, {\cal P}_{1} \,
{N_{1}}(x,p)  \; .   \label{eq:PM2}
\end{equation}
Taking the structures of the matrices (\ref{eq:Mas}) and
(\ref{eq:P1}),(\ref{eq:P2}) into account, one can easily obtain
\begin{equation}
{\cal P}_{1} \, \hat{M} = \mu^{-1} \, \hat{m} \,{\cal P}_{1} \; ,
\label{eq:PM3}
\end{equation}
\begin{equation}
{\cal P}_{2} \, \hat{M} = \mu \,{\cal P}_{2} \; ,  \label{eq:PM4}
\end{equation}
where the operator $\hat{m}$ is given by
\begin{equation}
{\hat{m}}(x,-\frac{i\hbar}{2} \frac{\partial}{\partial p}) = m^{2} - \xi
\hbar^{2} {\hat{R}}(x,-\frac{i\hbar}{2} \frac{\partial}{\partial p}) \; .
\label{eq:masop}
\end{equation}
Using (\ref{eq:PM3}) and (\ref{eq:PM4}) in (\ref{eq:PM1}) and (\ref{eq:PM2})
leads to the following equation:
\begin{equation}
\left( m^{2} - \xi \hbar^{2} {\hat{R}}(x,-\frac{i\hbar}{2}
\frac{\partial}{\partial p}) \right) \:
{\cal P}_{1} \, {N_{1}}(x,p) = \hat{K_{1}} \, \hat{K_{1}} \: {\cal P}_{1} \,
{N_{1}}(x,p)  \; .   \label{eq:PM5}
\end{equation}
Substituting the explicit form of the operator $\hat{K_{1}}$ from Eq.\
(\ref{eq:kin2}) into Eq.\ (\ref{eq:PM5}) and taking of the property
(\ref{eq:prop3}) into account, one can obtain the equation for the Wigner
function (\ref{eq:wigf}) of the scalar field by making use of the prescription
(\ref{eq:extrf}). The result is:
\begin{equation}
\left( m^{2} - \xi \hbar^{2} {\hat{R}}(x,-\frac{i\hbar}{2}
\frac{\partial}{\partial p}) \right) \:
{f}(x,p)
= \left( \hat{K}^{\alpha}
\, \hat{K}_{\alpha} + \frac{i\hbar}{2} \hat{R}_{1\alpha}
\, \hat{K}^{\alpha} \right) \: {f}(x,p) \; ,      \label{eq:kinf}
\end{equation}
where $\hat{K}_{\alpha}$ is the following operator:
\begin{equation}
\hat{K}_{\alpha} \: {f}(x,p) =
\left( \frac{i\hbar}{2} \hat{R}_{1\alpha}^{\nu} \tilde{D}_{\nu}
+\hat{R}_{2\alpha}^{\nu} \, p_{\nu} - \frac{e}{2}\, \hat{F}_{1\alpha} \right)\:
{f}(x,p)
 - \frac{e}{2}\, {f}(x,p)\,
\stackrel{\leftarrow}{\hat{F}}_{2\alpha} \; , \label{eq:Kalpha}
\end{equation}
and the operators $\hat{R}_{1\alpha}$ and $\hat{F}_{1,2\alpha}$ are defined by
Eqs.\ (\ref{eq:A12})--(\ref{eq:Ak}), replacing:
\begin{equation}
A_{s\alpha\nu} \longmapsto  \left\{ \begin{array}{ll}
 R_{\alpha\nu} & \mbox{for $\hat{R}_{1\alpha}$ } \\
 F_{\alpha\nu} & \mbox{for $\hat{F}_{1,2\alpha}$ }  \end{array}  \right. \; .
\label{eq:subst2}
\end{equation}
Eq.\ (\ref{eq:kinf}) contains, in fact, two equations. To see that, one should
remember that the Wigner function is an hermitian (matrix) function in view of
the definition (\ref{eq:wigf}). Separating the hermitian and the antihermitian
parts of Eq.\ (\ref{eq:kinf}), one will get two independent equations for the
only function ${f}(x,p)$, which have to be satisfied simultaneously. Here I
will only write out the explicit form of the classical terms and the first
quantum corrections of these equations.
 \\ The first quantum corrections due to the Yang--Mills field are multiplied
by $\hbar$ while the ones due to the gravitational field are multiplied by
$\hbar^{2}$. It is worth noting that to get the terms of the order of
$\hbar^{n}$ in both equations, one has to expand Eq.\ (\ref{eq:kinf}) up to the
terms of the order of $\hbar^{n+1}$, because the antihermitian part of it
includes the overall multiplyer $i\hbar$ (we suppose that quantum interference
effects have been quenched by averaging, so that the Wigner function itself is
independent on $\hbar$, see also the discussion after Eq.\
(\ref{eq:Ak})$^{\ref{kn:habib3}}$).
Then two equations to the single function will be consistent up to the terms of
the order of $\hbar^{n}$ (see the discussion below).\\  These two equations are
\begin{eqnarray}
\left( m^{2} - p^{\alpha}p_{\alpha} \right) \:
{f}(x,p) = \frac{ie\hbar}{4}\, p^{\nu} \frac{\partial}{\partial p_{\alpha}}
\left[ F_{\alpha\nu},{f}(x,p) \right]
- \frac{\hbar^{2}}{4}\, \tilde{D}^{\alpha} \tilde{D}_{\alpha} \: {f}(x,p) +
\nonumber \\
\mbox{} + \hbar^{2} \left( (\xi - \frac{1}{3} ) R - \frac{1}{12}
R_{\alpha\mu\beta\nu} p^{\mu} p^{\nu} \frac{\partial^{2}}{\partial p_{\alpha}
\partial p_{\beta}} - \frac{1}{4} R_{\mu\nu} p^{\mu} \frac{\partial}{\partial
p_{\nu}} \right) \: {f}(x,p) \; , \hspace*{.1in}    \label{eq:mashf}
\end{eqnarray}
\begin{eqnarray}
p^{\alpha} \tilde{D}_{\alpha} \, {f}(x,p) + \frac{e}{2} p^{\nu}
\frac{\partial}{\partial p_{\alpha}} \left\{ F_{\alpha\nu},{f}(x,p) \right\}
= \frac{ie\hbar}{8} \frac{\partial}{\partial p_{\alpha}} \left[
F_{\nu\alpha},\tilde{D}^{\nu} {f}(x,p) \right] -   \nonumber \\
\mbox{} - \frac{ie\hbar}{8} p^{\nu} \frac{\partial^{2}}{\partial p_{\alpha}
\partial p_{\beta}} \left[ F_{\nu\alpha ; \beta},{f}(x,p) \right]
 - \frac{ie^{2}\hbar}{16} \frac{\partial^{2}}{\partial p_{\alpha} \partial
p_{\beta}} \left[ F_{\alpha\nu} F_{\beta .\ }^{\nu} , {f}(x,p) \right] +
\nonumber \\
\mbox{} + \hbar^{2} \left(
 \frac{1}{6} R^{\nu}_{\beta\mu\alpha} p^{\mu} \frac{\partial^{2}}{\partial
p_{\alpha} \partial p_{\beta}} \tilde{D}_{\nu} -\frac{1}{24}
R_{\alpha\mu\beta\nu ; \sigma} p^{\mu} p^{\nu} \frac{\partial^{3}}{\partial
p_{\alpha} \partial p_{\beta} \partial p_{\sigma}} +  \right. \hspace*{.2in}
\nonumber \\
\left.
\mbox{} + \frac{1}{12} R_{\alpha}^{\nu} \frac{\partial}{\partial p_{\alpha}}
\tilde{D}_{\nu}
 - \frac{1}{24} R_{\alpha\beta ; \nu} p^{\nu} \frac{\partial^{2}}{\partial
p_{\alpha} \partial p_{\beta}} + \frac{1}{2} ( \xi - \frac{1}{4} ) R_{; \alpha}
\frac{\partial}{\partial p_{\alpha}} \right)\: {f}(x,p) \; .  \label{eq:tranf}
\end{eqnarray}
The $[,]$--bracket stand here for the commutator and the $\{,\}$--bracket --
for the anticommutator of two matrices in the gauge group's representation
space. Remember also that the operator $\tilde{D}_{\alpha}$ is defined as
following:
\begin{equation}
\tilde{D}_{\alpha} \, {f}(x,p) = \left( \partial
_{\alpha} + \Gamma_{\alpha\beta}^{\gamma}\: p_{\gamma}\:
\frac{\partial}{\partial p_{\beta}} \right)\: {f}(x,p) + \frac{ie}{\hbar}
\left[ A_{\alpha}, {f}(x,p) \right] \; .  \label{eq:dpf}
\end{equation}
Eqs.\ (\ref{eq:mashf}) and (\ref{eq:tranf}) were first obtained by
Winter$^{\ref{kn:winter1},\ref{kn:winter2}}$ through a very different method.
The first equation gives the quantum corrections to the mass--shell constraint
of classical particles (\ref{eq:mass}) and the second equation is the quantum
Liouville--Vlasov equation.
These two equations must be satisfied simultaneously. It could lead to
inconsistency with the classical limit if we were not able to reduce them to a
single equation. \\ Indeed, in the naive classical limit ($\hbar \rightarrow
0$) the Wigner function satisfies two classical equations:
\begin{equation}
\Omega \: {f}(x,p) = 0  \; ,  \label{eq:mas0}
\end{equation}
\begin{equation}
\hat{{\cal L}} \: {f}(x,p) = 0 \; ,  \label{eq:kin0}
\end{equation}
where
\begin{equation}
\Omega = p^{\alpha} p_{\alpha} - m^{2}
\end{equation}
and $\hat{{\cal L}}$ is the classical Liouville operator, which acts on the
left hand side of Eq.\ (\ref{eq:tranf}) ( it coincides with the left hand side
of Eq.\ (\ref{eq:vla}) if the gauge group is ${U}(1)$ ). One can easily check
that the operator $\hat{{\cal L}}$ annihilates $\Omega$ and therefore the
function
\begin{equation}
{f_{cl}}(x,p) = {F_{cl}}(x,p) \, {\delta}(\Omega)   \label{eq:Fcl}
\end{equation}
will satisfy Eq.\ (\ref{eq:kin0}) if ${F_{cl}}(x,p)$ satisfies the same
equation but on the mass--shell only. It means that the following equation
holds:
\begin{equation}
\hat{{\cal L}} \: {F_{cl}}(x,p) = \Omega \, {\Delta}(x,p) \; ,  \label{eq:Lmas}
\end{equation}
where ${\Delta}(x,p)$ is an arbitrary function, which is nonsingular on the
mass--shell.\\
\hspace*{.5in} We see that in the classical limit one is able to reduce two
equations for the Wigner function (which is a distributed function in the
relativistic case) to a single equation which coincides with the Vlasov
equation for the 7--dimensional distribution function on the mass--shell ( more
exactly, the sum of the distribution functions for particles and antiparticles,
compare with Ref.\ \ref{kn:degroot} ).
The same situation can be expected in each order in $\hbar$, because otherwise
one wouldn't get the right semiclassical limit.\\  To show that Eqs.\
(\ref{eq:mashf}) and (\ref{eq:tranf}) lead to a single equation up to the terms
of the next adiabatic order, I will use the method of Ref.\ \ref{kn:hu}. Let me
look for a solution to Eg.\ (\ref{eq:mashf}) of the form:
\begin{equation}
{f}(x,p) = {F_{0}}(x,p) \, {\delta}(\Omega) +
{F_{1}}(x,p) \, {\delta'}(\Omega) +
{F_{2}}(x,p) \, {\delta''}(\Omega)  \; ,  \label{eq:F02}
\end{equation}
where ${F_{0}}(x,p)$ is the quantum corrected distribution function {\it on}
the mass--shell:
\begin{equation}
{F_{0}}(x,p) = {F_{cl}}(x,p) + {F_{qu}}(x,p) \; ,   \label{eq:F0}
\end{equation}
and ${F_{1}}(x,p)$ and ${F_{2}}(x,p)$ have purely quantum origin. It is worth
noting that if one has been interested in all terms in the adiabatic expansion
then, as it is mentioned in Ref.\ \ref{kn:hu}, one should include the term
$\hbar (\xi - 1/6){R}(x)$ in $\Omega$ to get an improved asymptotic
approximation for the Wigner function. But as, for the moment, we are only
interested in the lowest adiabatic order, it is more convenient to consider
that this term also contributes to the {\em off} mass--shell part of the Wigner
function.\\
Using now the property:
\begin{equation}
\Omega \, {\delta^{(n)}}(\Omega) = - n \, {\delta^{(n-1)}}(\Omega) \; ,
\label{eq:Omn}
\end{equation}
and treating all derivatives of the $\delta$--function as linearly independent,
one can get easily from Eq.\ (\ref{eq:mashf}) the first quantum corrections to
the classical distribution function {\it off} the mass--shell:
\begin{equation}
{F_{1}}(x,p) = \hat{\Pi} \, {F_{cl}}(x,p)  \; ,    \label{eq:F1}
\end{equation}
\begin{equation}
{F_{2}}(x,p) = \frac{1}{2} \left[ \hat{\Pi} , \Omega \right] \, {F_{cl}}(x,p) =
 -\frac{\hbar^{2}}{3} R_{\mu\nu} p^{\mu} p^{\nu} {F_{cl}}(x,p) \; ,
\label{eq:F2}
\end{equation}
where $\hat{\Pi}$ is the operator acting on the right hand side of Eq.\
(\ref{eq:mashf}).\\
The substitution of the trial solution (\ref{eq:F02}) with
(\ref{eq:F0}),(\ref{eq:F1}) and (\ref{eq:F2}) into Eq.\ (\ref{eq:tranf}) gives,
after omitting the higher adiabatic order terms,
\begin{eqnarray}
{\delta}(\Omega) \, \hat{{\cal L}} \, \left( {F_{cl}}(x,p) + {F_{qu}}(x,p)
\right) + {\delta'}(\Omega) \, \hat{{\cal L}} \hat{\Pi} \, {F_{cl}}(x,p)
+ \frac{1}{2} {\delta''}(\Omega) \, \hat{{\cal L}} \left[ \hat{\Pi} , \Omega
\right] \, {F_{cl}}(x,p) = \nonumber \\
= {\delta}(\Omega) \, \hat{\Lambda} \, {F_{cl}}(x,p) +
{\delta'}(\Omega) \, \left[ \hat{\Lambda} , \Omega \right] \, {F_{cl}}(x,p) +
\frac{1}{2} {\delta''}(\Omega) \, \left[ \left[ \hat{\Lambda} , \Omega \right]
, \Omega \right] \, {F_{cl}}(x,p) \; , \hspace*{.1in}
 \label{eq:delts}
\end{eqnarray}
where $\hat{\Lambda}$ is the operator acting on the right hand side of Eq.\
(\ref{eq:tranf}) (the operators $\hat{\Pi}$ and $\hat{\Lambda}$ are written out
explicitly in ~\ref{sec-D}).\\
Eq.\ (\ref{eq:delts}) seems to give rise to inconsistency of the method, but
hopefully, the structures of the operators in Eqs.\ (\ref{eq:mashf}) and
(\ref{eq:tranf}) are so fine tuned that it leads to the cancelling of the terms
which multiply the derivatives of $\delta$--function.
Indeed, as it is shown in ~\ref{sec-D}, the operators $\hat{\Pi}$ and
$\hat{\Lambda}$ satisfy the following identity:
\begin{equation}
\left[ \hat{{\cal L}} , \hat{\Pi} \right] = \left[ \hat{\Lambda} , \Omega
\right] \; .  \label{eq:idPL}
\end{equation}
Using also the fact that the operator $\hat{{\cal L}}$ annihilates $\Omega$
and, therefore,
\begin{equation}
\left[ \hat{{\cal L}} , \left[ \hat{\Pi} , \Omega \right] \right] = \left[
\left[ \hat{{\cal L}} , \hat{\Pi} \right] , \Omega \right] \; ,
\label{eq:idPL2}
\end{equation}
one can transform Eq.\ (\ref{eq:delts}) into:
\begin{eqnarray}
{\delta}(\Omega) \, \hat{{\cal L}} \, \left( {F_{cl}}(x,p) + {F_{qu}}(x,p)
\right) + {\delta'}(\Omega) \, \hat{\Pi} \hat{{\cal L}} \, {F_{cl}}(x,p) +
\nonumber \\
\mbox{} + \frac{1}{2} {\delta''}(\Omega) \, \left[ \hat{\Pi} , \Omega \right]
\hat{{\cal L}} \, {F_{cl}}(x,p) =
 {\delta}(\Omega) \, \hat{\Lambda} \, {F_{cl}}(x,p) \; . \hspace*{.2in}
\label{eq:delts2}
\end{eqnarray}
Remember now that the classical distribution function ${F_{cl}}(x,p)$ satisfies
Eq.\ (\ref{eq:Lmas}). This leads, after using the property (\ref{eq:Omn}), to
the equation for evaluating the first quantum corrections to the classical
distribution function {\it on} the mass--shell:
\begin{equation}
{\delta}(\Omega) \, \hat{{\cal L}} \, {F_{qu}}(x,p)  =
 {\delta}(\Omega) \, \left( \hat{\Lambda} \, {F_{cl}}(x,p)
+ \hat{\Pi} \, {\Delta}(x,p) \right) \; .   \label{eq:Fqu}
\end{equation}
Eqs.\ (\ref{eq:F02}),(\ref{eq:F0}),(\ref{eq:F1}),(\ref{eq:F2}) and
(\ref{eq:Fqu}) solve, in principle, the task of finding the lowest order
quantum corrections to the classical distribution function (\ref{eq:Fcl}) of
scalar particles due to the coupling to external gravitational and Yang--Mills
fields. As it was emphasized in Ref.\ \ref{kn:hu}, some terms of the higher
order must be added, they give the right trace anomaly of the stress--energy
tensor.
\subsection{The Dirac field case}
\hspace*{.5in} An equation for the Wigner function of the Dirac field
(\ref{eq:wigN}) can be obtained by squaring Eq.\ (\ref{eq:com}) for $s=1/2$.
Remembering that in this case the operator $\hat{M}$ coincides with the field
mass $m$, one easily obtains:
\begin{equation}
m^{2} \, {N}(x,p) = \hat{K}_{1/2} \, \hat{K}_{1/2} \, {N}(x,p) \; .
\label{eq:eqN}
\end{equation}
Utilizing the algebra of the Dirac matrices (\ref{eq:comd}), one can deduce
from Eq.\ (\ref{eq:eqN}) the equations for the independent components of the
matrix Wigner function (\ref{eq:wigsc})--(\ref{eq:wigpsc}).
It happens that the equations form a coupled system of equations and other
components of the Wigner function (\ref{eq:wigN}) ( a vector and a pseudovector
functions ) defined by:
\begin{equation}
{A^{\alpha}}(x,p) =  Tr \left( \gamma^{\alpha} {N}(x,p) \right)   \; ,
\label{eq:wigvec}
\end{equation}
\begin{equation}
{B^{\alpha}}(x,p) = Tr \left( \gamma^{5} \gamma^{\alpha} {N}(x,p) \right) \; ,
    \label{eq:wigpvec}
\end{equation}
are expressed in terms of the functions (\ref{eq:wigsc})--(\ref{eq:wigpsc}) and
their derivatives only.
\\ The explicit form of the equations is very unwieldy in general and here I
will write out the equations with the first quantum corrections only ( of the
order of $\hbar$ ). The general form of the higher terms in expansions like
(\ref{eq:kinf})
for the Dirac field
can be found in Ref.\ \ref{kn:fonarevthes}.
It is only worth noting that quantum corrections of two types arise. The
corrections of the first type are similar to and coincide, up to the terms of
the order of $\hbar^{4}$, with the ones in Eq.\
(\ref{eq:mashf}),(\ref{eq:tranf}) when $\xi = 1/4$.
The corrections of the second type reflect the coupling of the spin and/or the
isospin of quantum particles to external gravitational and Yang--Mills fields
and they mix different components of the Wigner function
(\ref{eq:wigsc})--(\ref{eq:wigpsc}).
\\ As in the scalar field case, Eq.\ (\ref{eq:eqN}) leads to two systems of
equations: the constraint equations, which describe the quantum corrections to
the classical mass--shell constraint (\ref{eq:mass}), and the quantum transport
equations governing the evolution of the Wigner function.
These equations with the lowest order quantum corrections are
\begin{equation}
\left\{
\begin{array}{l}
 ( \Omega + \hat{\Pi} ) \, {A}(x,p) = \frac{e\hbar}{2} \left\{ F_{\mu\nu} ,
{A^{\mu\nu}}(x,p) \right\}   \\
 ( \Omega + \hat{\Pi} ) \, {B}(x,p) = \frac{e\hbar}{2} \left\{
\tilde{F}_{\mu\nu} , {A^{\mu\nu}}(x,p) \right\}   \\  \label{eq:mashN}
 ( \Omega + \hat{\Pi} ) \, {A_{\mu\nu}}(x,p) = \frac{e\hbar}{4} \left\{
F_{\mu\nu} , {A}(x,p) \right\} -
  \frac{e\hbar}{2} \left\{ \tilde{F}_{\mu\nu} , {B}(x,p) \right\} \; ,
\end{array}  \right.
\end{equation}
\begin{equation}
\left\{
\begin{array}{l}
 ( \hat{{\cal L}} - \hat{\Lambda} ) \, {A}(x,p) = - \hat{S}_{\mu\nu} \,
{A^{\mu\nu}}(x,p)  \\
( \hat{{\cal L}} - \hat{\Lambda} ) \, {B}(x,p) = - \hat{\tilde{S}}_{\mu\nu} \,
{A^{\mu\nu}}(x,p)  \\  \label{eq:kinN}
( \hat{{\cal L}} - \hat{\Lambda} ) \, {A_{\mu\nu}}(x,p) = - \frac{1}{2}
\hat{S}_{\mu\nu} \, {A}(x,p)  +
 \frac{1}{2} \hat{\tilde{S}}_{\mu\nu} \, {B}(x,p) \; , \end{array}
\right.   \hspace{.2in}
\end{equation}
where $\Omega$ and the operators $\hat{\Pi}$, $\hat{{\cal L}}$ and
$\hat{\Lambda}$ are defined as in the previous section (they are written out in
{}~\ref{sec-D}, only the terms of the order of $\hbar$ of $\hat{\Pi}$ and
$\hat{{\cal L}}$ are kept in Eqs.\ (\ref{eq:mashN}),(\ref{eq:kinN}) ) and
$\hat{S}_{\mu\nu}$ is the following operator:
\begin{equation}
\hat{S}_{\mu\nu} \, {A}(x,p) = \frac{ie}{2}\left[ F_{\mu\nu} , {A}(x,p) \right]
+ \frac{e\hbar}{2} R_{\beta\alpha\mu\nu} p^{\beta} \frac{\partial}{\partial
p_{\alpha}} {A}(x,p) + \frac{e\hbar}{4} \frac{\partial}{\partial p_{\alpha}}
\left\{ F_{\mu\nu ; \alpha} , {A}(x,p) \right\}  \; .   \label{eq:S}
\end{equation}
The tilde over tensors indicates dual tensors, for example,
$\tilde{F}_{\mu\nu} = \frac{1}{2} \sqrt{-{g}(x)} \,
\varepsilon_{\mu\nu\alpha\beta} \, F^{\alpha\beta}$.
\\ \hspace*{.5in} Eqs.\ (\ref{eq:mashN})--(\ref{eq:S}) were obtained for the
case of the ${U}(1)$--group ( an external electromagnetic field ) in Ref.\
\ref{kn:fonarev1}.
It is interesting to compare these equations with the heuristic equation
proposed by Israel$^{\ref{kn:israel}}$ for describing particles with internal
spin in external gravitational and electromagnetic fields. He introduces the
extended phase space by adding to it the spin angular momentum $s^{\mu\nu}$ of
particles, and the distribution function ${N}(x,p,s)$ is a function on this
space.
One can show that the transport equation to the function ${N}(x,p,s)$ leads,
after integrations over $s^{\mu\nu}$, to the equations
(\ref{eq:mashN})--(\ref{eq:S}), if one identifies the functions ${A}(x,p)$,
${B}(x,p)$ and ${A^{\mu\nu}}(x,p)$ with the corresponding moments of the
function ${N}(x,p,s)$ in the $s^{\mu\nu}$--space. The correspondence here is
quite similar to the one which arose in the case of colored particles in a
Yang--Mills field ( see, for instance, Ref.\ \ref{kn:heinz} ).
\section{Final remarks}
\setcounter{equation}{0}
\label{sec-final}
\hspace*{.5in} As a conclusion, I would like to mention a few further
developments related to the present work, which have not been considered in the
paper. (1) This paper didn't specify any regularization procedure as well as it
didn't discuss in detail the question of the renormalization of the statistical
averaged operators. This question is especially important when one wishes to
describe the particles creation due to quantum interactions or external fields.
This process could be described in kinetic theory, possibly by a source--term
in the Liouville equation  and could lead to Boltzmann entropy generation (
different approaches to defining a nonequilibrium entropy for quantum fields in
a cosmological setting have been proposed in Refs.\
\ref{kn:kandrup1}--\ref{kn:kandrup3}, see also Ref.\ \ref{kn:habib1} ).
(2) Specification of particles interactions will give an explicit expression
for the collision term in Eq.\ (\ref{eq:kin1}). As it is shown in Ref.\
\ref{kn:hu2}, a satisfactory approximation is provided only by the two--loop
order of the perturbation theory and higher. The theory becomes very involved
in this approximation and one needs some physical hypothesis to make further
progress.
(3) In Ref. \ref{kn:pirk} the quantum kinetic equation for the Wigner function
of a scalar field has been solved explicitly in the lowest adiabatic order in a
Robertson--Walker space--time, in Ref.\ \ref{kn:fonarev3} -- in a class of
manifolds admitting a Killing vector ( for instance, static space--times ) and
in Ref.\ \ref{kn:fonarevunp} -- a conformal Killing vector ( the later class
includes, in particular, a Robertson--Walker space--times ). Ref.\
\ref{kn:fonarevthes} considers also a few other physical systems permitting an
explicit analysis. In each case the quantum--curvature corrections to the
local--equilibrium distribution are expressed by
local geometrical quantities and
a few momentum derivatives of the classical distribution function.
This fact extremely simplifies the Einstein--Vlasov problem, reducing it to an
analysis of differential equations of a finite order ( such analysis by using
the moments method was carried out for the Friedmann cosmology in Ref.\
\ref{kn:messer}).
A natural question arises: is this property general or is it only a lucky
chance and how do space--time symmetries show themselves in the Wigner
function? The better understanding of the structure of the relativistic phase
space could possibly clarify this tie.
\section*{Acknowledgements}
\hspace*{.5in} It is a pleasure to express my indebtedness to Professor Yu. G.
Ignat'ev who stimulated my interest in kinetic theory. Valuable discussions
with the participants of the seminar on general--relativistic statistics and
cosmology of the Kazan Teachers Training Institute helped me to shape the
present work. I am also grateful to Professor G. Horwitz for helpful
discussions and a careful reading of the manuscript. I would like to thank
Professor L. P. Horwitz for fruitful discussions of the work.
\appendix
\renewcommand{\thesection}{Appendix \Alph{section}}
\renewcommand{\theequation}{\Alph{section}\arabic{equation}}
\section{ The Wigner function as expressed on the base space}
\setcounter{equation}{0}
\hspace*{.5in} The definition of a covariant and gauge invariant Wigner
function (\ref{eq:wigf}) is not very useful if one wishes to compute it
explicitly or explore its ultraviolet behavior, because the fields
\label{sec-A}
${{\bf \Phi}}(x,-y)$ and
${{\bf \Phi}^{\dagger}}(x,y)$  have been defined on the tangent space rather
than on the physical space--time.
Fortunately, one can express these fields in terms of fields on the base space,
though in general the expressions also include infinite series in the
derivatives.
\\ Consider first the field
${{\bf \Phi}}(x,-y)$. Let me rewrite expression (\ref{eq:Phi1}) in the
following form:
\begin{equation}
{{\bf \Phi}}(x,-y) = \left( \exp(-y^{\alpha} \hat{\tilde{\nabla}}_{\alpha}) \,
\exp(y^{\alpha} \partial_{\alpha}) \right) \: \left(
\exp(-y^{\alpha} \partial_{\alpha}) \,
{{\bf \varphi}}(x)  \right) \; .     \label{eq:rePhi1}
\end{equation}
The second cofactor in (\ref{eq:rePhi1}) is obviously
${{\bf \varphi}}(x-y)$
for any numerous value of $y^{\alpha}$ in each fixed coordinate system. The
first cofactor is in general a differential operator acting to
${{\bf \varphi}}(x-y)$.
It can be written more explicitly by using the Campbell--Baker--Hausdorff
formula ( see, for instance, Ref.\ \ref{kn:serre}):
\begin{equation}
\exp(\hat{A}) \, \exp(\hat{B}) = \exp \left( \sum_{n=1}^{\cal 1} \hat{C}_{n}
\right) \; ,  \label{eq:CBH}
\end{equation}
with $\hat{C}_{n}$ being found by the sequence:
\begin{eqnarray}
\hat{C}_{1} = \hat{A} + \hat{B} \; , \hspace*{2in} \nonumber \\
\hat{C}_{2} = \frac{1}{2} \left[ \hat{A} , \hat{B} \right] \; , \hspace*{1.9in}
\nonumber \\ \label{eq:seqC}
\hat{C}_{n} = \frac{1}{n!\ } \left[ \right. \overbrace{\hat{A} , \left[ \ldots
, \left[ \hat{A} \right. \right. }^{n-1} ,\left. \left. \hat{B} \right] \ldots
\right] \left. \right] - \hspace*{1in}  \nonumber \\
\mbox{} - \frac{1}{n} \sum_{m=2}^{n-1} \frac{1}{m!\ } \sum_{k_{1}} \cdots
\sum_{k_{m}}
k_{1} \delta_{n , k_{1} + \ldots + k_{m} } \: \left[ \hat{C}_{k_{m}} , \left[
\ldots , \left[ \hat{C}_{k_{2}} , \hat{C}_{k_{1}} \right] \ldots \right]
\right] \; .
\end{eqnarray}
If one puts first,
$\hat{B} =\- y^{\alpha} \hat{\nabla}_{\alpha}$ and
 $\hat{A} =\- - y^{\alpha}
\hat{\tilde{\nabla}}_{\alpha} =\- - \hat{B} -\- \frac{ie}{\hbar} y^{\alpha}
{A_{\alpha}}(x)$,
then it follows from Eqs.\ (\ref{eq:CBH}),(\ref{eq:seqC}) that
\begin{eqnarray}
{Z}(x,-y) = \exp(-y^{\alpha} \hat{\tilde{\nabla}}_{\alpha}) \,
\exp(y^{\alpha} \hat{\nabla}_{\alpha})  = \hspace*{1in}  \nonumber \\
= \exp \left( \frac{ie}{\hbar} \sum_{n=0}^{{\cal 1}} \frac{(-1)^{n+1}}{(n+1)!\
}
y^{\alpha_{0}} \cdots y^{\alpha_{n}} A_{\alpha_{0} ; \alpha_{1} \ldots
\alpha_{n}} -
 \frac{e^{2}}{12\hbar^{2}} y^{\alpha} y^{\beta} y^{\sigma} \left[ A_{\alpha ;
\beta} , A_{\sigma} \right] + \ldots \right) \; . \label{eq:Z1}
\end{eqnarray}
This expression can be written in a compact form by using Feynman's path
ordering operator $P$:
\begin{equation}
{Z}(x,-y) = P \; \exp \left( - \frac{ie}{\hbar} \int_{-1}^{1} dt\,
\sum_{n=0}^{{\cal 1}} \frac{t^{n}}{n!\ } y^{\alpha^{0}} \cdots y^{\alpha^{n}}
A_{\alpha_{0} : \alpha_{1} \ldots \alpha_{n}} \right) \; . \label{eq:Zcom}
\end{equation}
In the same way, with $\hat{B} = y^{\alpha} \partial_{\alpha}$ and
$\hat{A} = - y^{\alpha} \hat{\nabla}_{\alpha} = - \hat{B} +
\Gamma_{\alpha\sigma}^{\beta} y^{\alpha} y^{\sigma} \frac{\partial}{\partial
y^{\beta}}$ Eqs.\ (\ref{eq:CBH}),(\ref{eq:seqC}) give:
\begin{eqnarray}
{\hat{{\cal P}}}(x,-y) =
 \exp(-y^{\alpha} \hat{\nabla}_{\alpha}) \,
\exp(y^{\alpha} \partial_{\alpha}) = \hspace*{1in}  \nonumber \\
 = \exp \left( y^{\alpha} y^{\beta} \Gamma_{\alpha\beta}^{\sigma} \left(
\frac{1}{2}
\partial_{\sigma} + \frac{\partial}{\partial y^{\sigma}} \right) - y^{\alpha}
y^{\beta} y^{\gamma} \Gamma_{\alpha\beta ; \gamma}^{\sigma} \left( \frac{1}{3}
\partial_{\sigma} + \frac{1}{2} \frac{\partial}{\partial y^{\sigma}} \right) +
\ldots \right)  \; .   \label{eq:calP}
\end{eqnarray}
If one uses the Riemann normal coordinates$^{\ref{kn:kobayashi}}$ centered at
point $x^{\alpha}$ (RNC), the operator ${\hat{{\cal P}}}(x,-y)$ vanishes. Then
the field
${{\bf \Phi}}(x,-y)$
differs from
${{\bf \varphi}}(x-y)$ only by a phase, which is (\ref{eq:Zcom}). \\
 In the same way, the field
${{\bf \Phi}^{\dagger}}(x,y)$ in RNC equals to
${Z^{\dagger}}(x,y)
{{\bf \phi}^{\dagger}}(x+y)$  and
the Wigner function (\ref{eq:wigf}) becomes
\begin{equation}
{f}(x,p)_{|RNC} = (\pi\hbar)^{-4}\: \int
d^{4}\,y\:e^{-2iy^{\alpha}p_{\alpha}/\hbar}\:
{Z}(x,-y)
\langle {{\bf \phi}}(x-y){{\bf \phi}^{\dagger}}(x+y)\rangle
{Z^{\dagger}}(x,y)
\label{eq:wigRNC}
\end{equation}
 In absence of an external Yang--Mills field, ${Z}(x,-y) = 1$ and the function
(\ref{eq:wigRNC})  coincides with that defined in Ref.\ \ref{kn:hu} ( in RNC )
within a factor equal to the square root of the Van Vleck--Morett determinant,
or the one defined in Ref.\ \ref{kn:winter2}. In other coordinate systems the
Wigner function can be calculated by using Eq.\ (\ref{eq:calP}) ( see also the
discussion in Ref.\ \ref{kn:hu} ). \\
In another limiting case, when the space--time is flat, the tangent space
coincides with Minkovski space.
In this case the function (\ref{eq:Zcom}) takes the familiar form:
\begin{equation}
{Z}(x,-y) = P \; \exp \left( - \frac{ie}{\hbar} \int_{0}^{1} dt\, y^{\alpha}
{A_{\alpha}}(x - ty) \right) \; ,  \label{eq:Zfl}
\end{equation}
and the gauge--invariant Wigner function (\ref{eq:wigRNC}) coincides with that
defined in Ref.\ \ref{kn:winter1} or in Ref.\ \ref{kn:heinz}, in the Dirac
field case (for the case of the gauge group being ${U}(1)$, see for instance
Ref.\ \ref{kn:akhiezer}).
\\ \hspace*{.5in} Thus, our definition of the Wigner function in curved
space--time is locally equivalent to those of Refs.\
\ref{kn:winter2},\ref{kn:hu}. But for nontrivial topologies, the different
definitions of the Wigner function for a system in a highly coherent state
could lead to globally different results.
\\ The same analysis can be carried out for the Wigner function of a Dirac
field (\ref{eq:wigN}).
\section{
Useful identities}
\setcounter{equation}{0}
\hspace*{.5in} In this Appendix I will prove a few useful identities, which
have been used in deriving the equation for the unified Wigner function
(\ref{eq:kin2}).\\
\label{sec-B}
Let $\hat{A}$ and $\hat{B}$ be two arbitrary operators. Then the following well
known identity holds:
\begin{equation}
\left[ \hat{A} , e^{\hat{B}} \right] = - \sum_{n=1}^{{\cal 1}} \frac{1}{n!\ }
\left[ \right. \overbrace{ \hat{B} , \left[ \cdots , \left[ \hat{B} \right.
\right. }^{n} , \left. \left. \hat{A} \right] \cdots \right] \left.
\right] \; e^{\hat{B}}  \; .       \label{eq:A21}
\end{equation}
Putting $\hat{A}= \hat{\tilde{\nabla}}_{\alpha}$ and
$\hat{B}= - y^{\nu} \hat{\tilde{\nabla}}_{\nu}$ leads to
\begin{equation}
 \hat{\tilde{\nabla}}_{\alpha} \, {{\bf U}}(x,-y) =
 e^{-y^{\nu} \hat{\tilde{\nabla}}_{\nu}} \, \hat{\tilde{\nabla}}_{\alpha} \,
{{\bf u}}(x) -
{\hat{H}_{\alpha}}(x,-y)
{{\bf U}}(x,-y) \; ,       \label{eq:Halpha}
\end{equation}
with
\begin{equation}
{\hat{H}_{\alpha}}(x,-y) =
\sum_{n=1}^{\cal 1} \frac{(-1)^{n}}{n!\ } y^{\nu_{1}} \ldots y^{\nu_{n}}
\left[ \hat{\tilde{\nabla}}_{\nu_{1}} ,
 \left[ \cdots , \left[
 \hat{\tilde{\nabla}}_{\nu_{n}} ,
 \hat{\tilde{\nabla}}_{\alpha} \right] \cdots \right] \right] \; .
\label{eq:H}
\end{equation}
The field
${{\bf U}}(x,-y)$ is defined by the same formula as in (\ref{eq:Phi1}):
\begin{equation}
{{\bf U}}(x,-y) = \exp(-y^{\alpha} \hat{\tilde{\nabla}}_{\alpha}) \, {{\bf
u}}(x)   \; .
\end{equation}
If one puts $\hat{A} = {M}(x)$, then (\ref{eq:A21}) with the same $\hat{B}$
gives the following identity:
\begin{equation}
 e^{-y^{\nu} \hat{\tilde{\nabla}}_{\nu}} \, {M}(x) {{\bf u}}(x) = {M}(x,-y)
{{\bf U}}(x,-y) \; ,   \label{eq:comM}
\end{equation}
with
\begin{equation}
{M}(x,-y) = {M}(x) +
\sum_{n=1}^{\cal 1} \frac{(-1)^{n}}{n!\ } y^{\nu_{1}} \ldots y^{\nu_{n}} \,
\tilde{\nabla}_{\nu{1}} \cdots \tilde{\nabla}_{\nu_{n}} {M}(x) \; .
\label{eq:Mxy}
\end{equation}
The next identity follows from the formula for the left derivative of the
exponent of an operator function ${\hat{B}}(t)$:
\begin{equation}
 \frac{d}{dt} e^{{\hat{B}}(t)} = \left(
 \frac{d{\hat{B}}(t)}{dt} +
 \sum_{n=1}^{\cal 1} \frac{1}{(n+1)!\ }
\left[ \right. \overbrace{ {\hat{B}}(t) , \left[ \cdots , \left[ {\hat{B}}(t)
\right. \right. }^{n}
 ,\frac{d{\hat{B}}(t)}{dt} \left. \left.
\right] \cdots \right] \left. \right] \right)
 \, e^{{\hat{B}}(t)} \; .   \label{eq:leftd}
\end{equation}
Substituting ${\hat{B}}(y^{\alpha}) = -
 y^{\alpha} \hat{\tilde{\nabla}}_{\alpha}
 - \sum_{\nu \neq \alpha} y^{\nu} \hat{\tilde{\nabla}}_{\nu}$ (with no
summation over $\alpha$) into (\ref{eq:leftd}) gives the identity:
\begin{equation}
 \frac{\partial}{\partial y^{\alpha}} \, {{\bf U}}(x,-y) =
 - \hat{\tilde{\nabla}}_{\alpha} \, {{\bf U}}(x,-y) -
 {\hat{G}_{\alpha}}(x,-y){{\bf U}}(x,-y) \; , \label{eq:dyU}
\end{equation}
where
\begin{equation}
 {\hat{G}_{\alpha}}(x,-y) =
\sum_{n=1}^{\cal 1} \frac{(-1)^{n}}{(n+1)!\ } y^{\nu_{1}} \ldots y^{\nu_{n}}
\left[
 \hat{\tilde{\nabla}}_{\nu_{1}} ,
 \left[ \cdots , \left[
 \hat{\tilde{\nabla}}_{\nu_{n}} ,
 \hat{\tilde{\nabla}}_{\alpha} \right] \cdots \right] \right] \; .
\label{eq:Gal}
\end{equation}
The same identity for the conjugate field is obtained by reversing the sign of
$y^{\alpha}$.
\section{
 Evaluation of $G_{s\alpha}(x,y)$}
\setcounter{equation}{0}
\hspace*{.5in} In this Appendix I evaluate expression (\ref{eq:G}) with the
operators $\hat{H}$ and $\hat{G}$ being defined in (\ref{eq:H}),(\ref{eq:Gal}).
I will omit here the subscript $s$ for convenience. \\
\label{sec-C}
Let me first prove the following recursive formula ($n \geq 2$):
\begin{equation}
G_{\alpha}^{(n)} = - A_{\alpha}^{(n-1)} \, {{\bf U}}(x,-y)
 + R_{\alpha}^{(n) \nu} \, \frac{\partial}{\partial y^{\nu}} \, {{\bf U}}(x,-y)
- \sum_{k=1}^{n-2} C_{k}^{n-2} \, R_{\alpha}^{(n-k) \nu} \, G_{\nu}^{(k)} \; ,
\label{eq:recf}
\end{equation}
where $C_{k}^{n} = n!\ /(n-k)!\ k!\ $ are the binomial coefficients and
\begin{equation}
\left\{ \begin{array}{l}
G_{\alpha}^{(n+1)} =
 y^{\nu_{1}} \ldots y^{\nu_{n}} \left[
 \hat{\tilde{\nabla}}_{\nu_{1}} ,
 \left[ \cdots , \left[
 \hat{\tilde{\nabla}}_{\nu_{n}} ,
 \hat{\tilde{\nabla}}_{\alpha} \right] \cdots \right] \right]\, {{\bf U}}(x,-y)
  \hspace{.1in} \mbox{if $n \geq 1$}  \\
G_{\alpha}^{(1)} =
 \hat{\tilde{\nabla}}_{\alpha} \, {{\bf U}}(x,-y) \; , \end{array}
\right. \label{eq:Gn}
\end{equation}
\begin{equation}
\left\{ \begin{array}{l}
R_{\alpha}^{(n) \beta} =
 y^{\nu_{1}} \ldots y^{\nu_{n}} R^{\beta}_{\nu_{1} \alpha \nu_{2} ; \nu_{3}
\ldots \nu_{n} }
   \hspace{.1in} \mbox{if $n \geq 3$}  \\
R_{\alpha}^{(2) \beta} =
 y^{\nu_{1}} y^{\nu_{2}} R^{\beta}_{\nu_{1} \alpha \nu_{2}} \; ,
  \end{array}  \right.   \label{eq:Rn}
\end{equation}
\begin{equation}
\left\{ \begin{array}{l}
A_{\alpha}^{(n)} =
 y^{\nu_{1}} \ldots y^{\nu_{n}} A_{\alpha \nu_{1} ; \nu_{2} \ldots \nu_{n} }
   \hspace{.1in} \mbox{if $n \geq 2$}  \\
A_{\alpha}^{(1)} =
 y^{\nu} A_{\alpha \nu} \; ,
 \end{array}  \right.   \label{eq:An}
\end{equation}
with $A_{\alpha\nu}$ being defined by Eq.\ (\ref{eq:A}). \\
For $n=2$, the last term in Eq.\ (\ref{eq:recf}) disappears and the validity of
this formula follows from the definition (\ref{eq:dy}) and Eq.\
(\ref{eq:comu1}). Suppose that Eq.\ (\ref{eq:recf}) holds up to some fixed $n$.
Then, for $n+1$, one can get the equality:
\begin{eqnarray}
G_{\alpha}^{(n+1)} =
  - A_{\alpha}^{(n)} \, {{\bf U}}(x,-y)
 + R_{\alpha}^{(n+1) \nu} \, \frac{\partial}{\partial y^{\nu}} \, {{\bf
U}}(x,-y) - \hspace*{.5in} \nonumber \\
\mbox{} - R_{\alpha}^{(n) \nu} \,
 \hat{\tilde{\nabla}}_{\beta} \, {{\bf U}}(x,-y)
  - \sum_{k=1}^{n-2} C_{k}^{n-2} \left(
R_{\alpha}^{(n-k+1) \nu} \, G_{\nu}^{(k)} +
R_{\alpha}^{(n-k) \nu} \, G_{\nu}^{(k+1)} \right) \; . \label{eq:recnext}
\end{eqnarray}
The proof will be completed if one takes into account the properties of the
binomial coefficients:
\begin{equation}
\begin{array}{l}
C_{k}^{n} + C_{k-1}^{n} = C_{k}^{n+1}    \\
C_{n}^{n} = 1  \; .
\end{array}
\end{equation}
Let me now define the series:
\begin{equation}
G_{\alpha}[g] = \sum_{n=2}^{\cal 1} g_{n} G_{\alpha}^{(n)} \; ,
\label{eq:Ggdef}
\end{equation}
with arbitrary $g_{n}$'s depending on $n$.
A repeated use of the formula (\ref{eq:recf}) gives:
\begin{eqnarray}
G_{\alpha}[g] = - \sum_{k=2}^{\cal 1} g_{k}
 \left( A_{\alpha}^{(k-1)}
- R_{\alpha}^{(k) \nu} \, \frac{\partial}{\partial y^{\nu}}  + (k-2)
R_{\alpha}^{(k-1) \nu}
 \hat{\tilde{\nabla}}_{\beta} \right) \, {{\bf U}}(x,-y) + \nonumber  \\
 \mbox{} + \sum_{n=2}^{\cal 1}\sum_{k_{1}=2}^{\cal 1}\ldots\sum_{k_{n}=2}^{\cal
1} (-1)^{n} \,
g_{s_{n}}\, C_{s_{1}}^{s_{2}-2} \cdots C_{s_{n-1}}^{s_{n}-2} \,
R_{\alpha}^{(k_{n})\beta_{n}} \ldots R_{\beta_{3}}^{(k_{2})\beta_{2}} \times
\hspace*{.2in} \nonumber \\
\times \left(
 A_{\beta_{2}}^{(k_{1}-1)}
- R_{\beta_{2}}^{(k_{1}) \nu} \, \frac{\partial}{\partial y^{\nu}}  + (k_{1} -
2) R_{\beta_{2}}^{(k_{1}-1) \nu}
 \hat{\tilde{\nabla}}_{\beta} \right) \, {{\bf U}}(x,-y) \; , \hspace*{.3in}
\label{eq:Gg}
\end{eqnarray}
where
\begin{equation}
s_{m} = k_{1} + \ldots + k_{m} \; . \label{eq:sm}
\end{equation}
By putting $g_{n}=(-1)^{n}/n!\ $ and $g_{n}= (-1)^{n}/(n+1)!\ $, one will get
expressions for the acting of the operators (\ref{eq:H}) and (\ref{eq:Gal}) to
${{\bf U}}(x,-y)$ and then evaluate the function ${G_{\alpha}}(x,y)$,
(\ref{eq:G}). \\ Unfortunately, the expression for ${G_{\alpha}}(x,y)$ includes
terms with \\
 $\left( \frac{\partial}{\partial y^{\alpha}} \, {{\bf U}}(x,-y) \right)
 {\bar{{\bf U}}}(x,y) $ or
 $\left( \hat{\tilde{\nabla}}_{\alpha} \, {{\bf U}}(x,-y) \right)
{\bar{{\bf U}}}(x,y) $.
In order to obtain
 an expression including only operators acting to
 ${{\bf U}}(x,-y)
 {\bar{{\bf U}}}(x,y) $ , one should use the identity (\ref{eq:dyU}) for field
${\bf U}$ as well as for the conjugate field. It is easy to get:
\begin{eqnarray}
 \left( \frac{\partial}{\partial y^{\alpha}} \, {{\bf U}}(x,-y) \right)
 {\bar{{\bf U}}}(x,y)
 = \frac{1}{2} \frac{\partial}{\partial y^{\alpha}} \left( {{\bf U}}(x,-y)
 {\bar{{\bf U}}}(x,y) \right) -
 \frac{1}{2} \hat{\tilde{\nabla}}_{\alpha} \left( {{\bf U}}(x,-y)
 {\bar{{\bf U}}}(x,y) \right) - \nonumber \\
\mbox{} - \frac{1}{2} {{\bf U}}(x,-y) \left( {\hat{G}_{\alpha}}(x,y){\bar{{\bf
U}}}(x,y) \right)
- \frac{1}{2} \left( {\hat{G}_{\alpha}}(x,-y){{\bf U}}(x,-y) {\bar{{\bf
U}}}(x,y)  \right) \; , \hspace*{.4in} \label{eq:difid1}
\end{eqnarray}
\begin{eqnarray}
 \left( \hat{\tilde{\nabla}}_{\alpha} \, {{\bf U}}(x,-y) \right)
 {\bar{{\bf U}}}(x,y)
 = - \frac{1}{2} \frac{\partial}{\partial y^{\alpha}} \left( {{\bf U}}(x,-y)
 {\bar{{\bf U}}}(x,y) \right) +
 \frac{1}{2} \hat{\tilde{\nabla}}_{\alpha} \left( {{\bf U}}(x,-y)
 {\bar{{\bf U}}}(x,y) \right) +  \nonumber \\
\mbox{} + \frac{1}{2} {{\bf U}}(x,-y) \left( {\hat{G}_{\alpha}}(x,y){\bar{{\bf
U}}}(x,y) \right)
- \frac{1}{2} \left( {\hat{G}_{\alpha}}(x,-y){{\bf U}}(x,-y) {\bar{{\bf
U}}}(x,y)  \right) \; .  \hspace*{.4in} \label{eq:difid2}
\end{eqnarray}
A repeated use of the identity (\ref{eq:Gg}) gives then, after elaborate
calculations, the formal representation for the function ${G_{\alpha}}(x,y)$,
(\ref{eq:G}) as an infinite series in operators acting to the product
 $ {\bf U\bar{U}} = {{\bf U}}(x,-y)
 {\bar{{\bf U}}}(x,y) $:
\begin{equation}
{G_{\alpha}}(x,y) = \left( K_{1\alpha}^{\nu}
  \hat{\tilde{\nabla}}_{\nu} -
 K_{2\alpha}^{\nu}
  \frac{\partial}{\partial y^{\nu}} + A_{1\alpha} \right) \, {\bf U\bar{U}}
 + {\bf U\bar{U}} \, A_{2\alpha} \; ,     \label{eq:Gsery}
\end{equation}
where
\begin{equation}
K_{1,2\alpha}^{\nu} =  \sum_{k=2}^{\cal 1} K_{k\alpha}^{1,2\beta}
R_{\beta}^{(k)\nu} \; ,
\end{equation}
\begin{equation}
A_{1,2\alpha} = \sum_{k=2}^{\cal 1} K_{k\alpha}^{3,4\beta} A_{\beta}^{(k-1)} \;
,
\end{equation}
with
\begin{equation}
K_{k\alpha}^{i\beta} = \delta_{\alpha}^{\beta} C_{k}^{i} + \sum_{n=1}^{\cal
1}\sum_{k_{1}=2}^{\cal 1}\ldots\sum_{k_{n}=2}^{\cal 1} C_{kk_{1}\ldots
k_{n}}^{i} R_{\alpha}^{(k_{n})\beta_{n}}\ldots R_{\beta_{2}}^{(k_{1})\beta} \;
, \label{eq:Kiy}
\end{equation}
and $R_{\alpha}^{(k)\beta}$ and $A_{\alpha}^{(k)}$ being defined by
(\ref{eq:Rn}),(\ref{eq:An}). \\
The numerical coefficients $C_{k_{1}\ldots k_{n}}^{i}$ can explicitly be found
in the following way.
Let ${A}(t)$ be an operator function. Let's introduce the following generating
functionals:
\begin{equation}
K^{i}[A] =  \sum_{n=1}^{\cal 1}\sum_{k_{1}=2}^{\cal
1}\ldots\sum_{k_{n}=2}^{\cal 1} C_{k_{1}\ldots k_{n}}^{i} A^{(k_{1})} \ldots
A^{(k_{n})} \; , \label{eq:genf}
\end{equation}
where $A^{(k)} = d^{k} {A}(0)/dt^{k}$ are the derivatives of ${A}(t)$ at
$t=0$.\\
Then,
the coefficients $C_{k_{1}\ldots k_{n}}^{i}$ are found
if we define
 these functionals as follows:
\begin{equation}
K^{1}[A] = -1 + K \, B \; , \label{eq:K1A}
\end{equation}
\begin{equation}
K^{2}[A] = -1 + K \, C \; , \label{eq:K2A}
\end{equation}
\begin{equation}
K^{3}[A] = -2 \bar{C}' \, \bar{C}^{-1} + K \, C \,\bar{C}^{-1} \left( \bar{C}
-1 \right) \; , \label{eq:K3A}
\end{equation}
\begin{equation}
K^{4}[A] = - K \, \left( C -1 \right) \; , \label{eq:K4A}
\end{equation}
where
\begin{equation}
K = \left( 1 - \bar{B}' \, \bar{B}^{-1} + \bar{C}' \, \bar{C}^{-1} \right) \,
 \left( \frac{1}{2} B \, \bar{B}^{-1} +
 \frac{1}{2} C \, \bar{C}^{-1} \right)^{-1} \; .
  \label{eq:KA}
\end{equation}
Here $B={B}(1)$, $C={C}(1)$, $\bar{B}={B}(-1)$ and $\bar{C}={C}(-1)$, where
${B}(t)$ and ${C}(t)$ are the following functionals:
\begin{equation}
{B}(t) = 1 +  \sum_{n=1}^{\cal 1}\sum_{k_{1}=2}^{\cal
1}\ldots\sum_{k_{n}=2}^{\cal 1} (-1)^{n} t^{s_{n}} \, \prod_{i=1}^{n}
\frac{1}{(k_{i}-2)!\ s_{i} (s_{i}+1)} A^{(k_{i})} \; , \label{Bt1}
\end{equation}
\begin{equation}
{C}(t) = 1 +  \sum_{n=1}^{\cal 1}\sum_{k_{1}=2}^{\cal
1}\ldots\sum_{k_{n}=2}^{\cal 1} (-1)^{n} t^{s_{n}} \, \prod_{i=1}^{n}
\frac{1}{(k_{i}-2)!\ s_{i} (s_{i}-1)} A^{(k_{i})} \; . \label{Ct1}
\end{equation}
 $\bar{B}'$ and $\bar{C}'$ denote the derivatives at $t=-1$.
\\ The functionals
${B}(t)$ and ${C}(t)$ can be written in a compact form by using
Feynman's path ordering operator $P$:
\begin{equation}
{B}(t) = P \; \exp \left( \int_{0}^{t} dx\, {b}(x) \right) \; ,  \label{eq:Bt2}
\end{equation}
\begin{equation}
{C}(t) = P \; \exp \left( \int_{0}^{t} dx\, {c}(x) \right) \; ,  \label{eq:Ct2}
\end{equation}
with
\begin{equation}
{b}(x) =  x^{-2} \, \int_{0}^{x} dz\, z^{2} {A^{(2)}}(z) \; ,  \label{eq:bx}
\end{equation}
\begin{equation}
{c}(x) =   \int_{0}^{x} dz\, {A^{(2)}}(z) \; .  \label{eq:cx}
\end{equation}
In particular, the coefficients in Eq.\ (\ref{eq:Kiy}) needed only for
evaluating the linear terms with respect to the curvature tensor in Eq.\
(\ref{eq:kin2}) are
\begin{equation}
C_{k}^{1} = (-1)^{k} / k!\ + \left( 1 - (-1)^{k} \right) / (k+1)!\ \; ,
\label{eq:Ck1}
\end{equation}
\begin{equation}
C_{k}^{2} = (-1)^{k} / k!\ - \left( 1 + (-1)^{k} \right) / (k+1)!\ \; ,
\label{eq:Ck2}
\end{equation}
\begin{equation}
C_{k}^{3} = (-1)^{k} ( 2k - 1) / k!\ \; ,  \label{eq:Ck3}
\end{equation}
\begin{equation}
C_{k}^{4} =  1 / k!\  \; , \label{eq:Ck4}
\end{equation}
\begin{eqnarray}
C_{k_{1}k_{2}}^{3} = (-1)^{k_{1}+k_{2}+1} \frac{2(k_{1}+k_{2}) - 1}{k_{1}!\
(k_{2}-2)!\ (k_{1}+k_{2})(k_{1}+k_{2}-1) } + \hspace*{.1in} \nonumber \\
\mbox{} + (-1)^{k_{1}+k_{2}} \left(1+(-1)^{k_{2}} \right) \frac{1}{k_{1}!\
(k_{2}+1)!\ }+ (-1)^{k_{1}+k_{2}} \frac{2(k_{1}-1)}{k_{1}!\ k_{2}!\ } \; ,
\label{eq:Ckk3}
\end{eqnarray}
\begin{eqnarray}
C_{k_{1}k_{2}}^{4} =  \frac{1}{k_{1}!\ (k_{2}-2)!\ (k_{1}+k_{2})(k_{1}+k_{2}-1)
} + \nonumber \\
\mbox{} + \left(1+(-1)^{k_{2}} \right) \frac{k_{2}}{k_{1}!\ (k_{2}+1)!\ } \; .
\hspace*{.5in}
 \label{eq:Ckk4}
\end{eqnarray}
\section{
Check of the consistency of the mass--shell constraint}
\setcounter{equation}{0}
\hspace*{.5in} This \label{sec-D} Appendix proves the identity (\ref{eq:idPL}).
Let me explicitly write out the operators
 $\hat{{\cal L}} ,
 \hat{\Pi} ,
 \hat{\Lambda}$
and also $ \Omega$:
\begin{equation}
 \hat{{\cal L}}
{f}(x,p) =
p^{\alpha} \tilde{D}_{\alpha} \, {f}(x,p) + \frac{e}{2} p^{\nu}
\frac{\partial}{\partial p_{\alpha}} \left\{ F_{\alpha\nu},{f}(x,p) \right\} \;
, \label{eq:Lap}
\end{equation}
\begin{eqnarray}
 \hat{\Pi}
{f}(x,p) =
 \frac{ie\hbar}{4}\, p^{\nu} \frac{\partial}{\partial p_{\alpha}} \left[
F_{\alpha\nu},{f}(x,p) \right]
- \frac{\hbar^{2}}{4}\, \tilde{D}^{\alpha} \tilde{D}_{\alpha} \: {f}(x,p) +
\hspace*{.2in} \nonumber \\
\mbox{} + \hbar^{2} \left( (\xi - \frac{1}{3} ) R - \frac{1}{12}
R_{\alpha\mu\beta\nu} p^{\mu} p^{\nu} \frac{\partial^{2}}{\partial p_{\alpha}
\partial p_{\beta}} - \frac{1}{4} R_{\mu\nu} p^{\mu} \frac{\partial}{\partial
p_{\nu}} \right) \: {f}(x,p)  \; ,   \label{eq:Piap}
\end{eqnarray}
\begin{eqnarray}
 \hat{\Lambda}
{f}(x,p)
= \frac{ie\hbar}{8} \frac{\partial}{\partial p_{\alpha}} \left[
F_{\nu\alpha},\tilde{D}^{\nu} {f}(x,p) \right] - \hspace*{.5in} \nonumber   \\
\mbox{} - \frac{ie\hbar}{8} p^{\nu} \frac{\partial^{2}}{\partial p_{\alpha}
\partial p_{\beta}} \left[ F_{\nu\alpha ; \beta},{f}(x,p) \right]
 - \frac{ie^{2}\hbar}{16} \frac{\partial^{2}}{\partial p_{\alpha} \partial
p_{\beta}} \left[ F_{\alpha\nu} F_{\beta .\ }^{\nu} , {f}(x,p) \right] +
\nonumber \\
\mbox{} + \hbar^{2} \left(
 \frac{1}{6} R^{\nu}_{\beta\mu\alpha} p^{\mu} \frac{\partial^{2}}{\partial
p_{\alpha} \partial p_{\beta}} \tilde{D}_{\nu} -\frac{1}{24}
R_{\alpha\mu\beta\nu ; \sigma} p^{\mu} p^{\nu} \frac{\partial^{3}}{\partial
p_{\alpha} \partial p_{\beta} \partial p_{\sigma}} +  \right. \hspace*{.1in}
\nonumber \\
\left.
 \mbox{} + \frac{1}{12} R_{\alpha}^{\nu} \frac{\partial}{\partial p_{\alpha}}
\tilde{D}_{\nu}
- \frac{1}{24} R_{\alpha\beta ; \nu} p^{\nu} \frac{\partial^{2}}{\partial
p_{\alpha} \partial p_{\beta}} + \frac{1}{2} ( \xi - \frac{1}{4} ) R_{; \alpha}
\frac{\partial}{\partial p_{\alpha}} \right)\: {f}(x,p) \; ,  \label{eq:Lamap}
\end{eqnarray}
\begin{equation}
\Omega = p^{\alpha} p_{\alpha} - m^{2} \; . \label{eq:Omap}
\end{equation}
To prove the identity
(\ref{eq:idPL}), one needs an expression for the commutator of two operators
$\tilde{D}_{\alpha}$
defined by Eq.\ (\ref{eq:dpf}). It can be shown that
\begin{equation}
 \left[ \tilde{D}_{\alpha} ,
  \tilde{D}_{\beta} \right] =
\left[ \tilde{\nabla}_{\alpha} ,
\tilde{\nabla}_{\beta} \right] -
R_{\alpha\beta\mu\nu} p^{\nu} \frac{\partial}{\partial p_{\mu}} \; ,
\label{eq:comap}
\end{equation}
where $
 \tilde{\nabla}_{\alpha} $ is the covariant and gauge invariant derivative
operator.\\
Therefore (compare with the formula (\ref{eq:recf}) ),
\begin{equation}
 \left[ \tilde{D}_{\alpha} ,
  \tilde{D}_{\beta} \right]
{f}(x,p) =
 \frac{ie}{\hbar} \, \left[ F_{\alpha\beta} ,
{f}(x,p) \right] -
R_{\alpha\beta\mu\nu} p^{\nu} \frac{\partial}{\partial p_{\mu}}
{f}(x,p)
\label{eq:com2ap}
\end{equation}
and
\begin{eqnarray}
 p^{\alpha}
 \left[ \tilde{D}_{\alpha} ,
  \tilde{D}^{\nu}
  \tilde{D}_{\nu} \right]
{f}(x,p) =
 \frac{ie}{\hbar} \,
 p^{\alpha}
\left[ F_{\alpha\nu}^{; \nu} ,
{f}(x,p) \right] +
 \frac{ie}{\hbar} \,
 p^{\alpha}
\left[ F_{\alpha\nu} ,
  \tilde{D}^{\nu}
{f}(x,p) \right] + \hspace*{.2in} \nonumber \\
\mbox{} + \left( 2 R_{\alpha\mu\beta\nu} p^{\alpha} p^{\beta}
\frac{\partial}{\partial p_{\mu}}
  \tilde{D}^{\nu}
- R_{\alpha\nu} p^{\alpha}
  \tilde{D}^{\nu}
- R_{\alpha\nu ; \beta} p^{\alpha} p^{\beta} \frac{\partial}{\partial p_{\nu}}
+
 R_{\alpha\beta ; \nu} p^{\alpha} p^{\beta} \frac{\partial}{\partial p_{\nu}}
\right)
\, {f}(x,p) \; .
\end{eqnarray}
Then, by noting the following property:
\begin{equation}
\left\{ \hat{A} , \left[ \hat{A} , f \right] \right\} -
\left[ \hat{A} , \left\{ \hat{A} , f \right\} \right] = 0 \; ,
\end{equation}
one can prove after elaborate calculations that the identity (\ref{eq:idPL}) is
right up to the terms of the next adiabatic order.
\vspace{2\baselineskip}
\newcounter{refer}
\begin{list}{$^{\arabic{refer}}$}{\usecounter{refer}
\setlength{\rightmargin}{\leftmargin} \footnotesize}
\item \label{kn:winter2} J. Winter, Phys. Rev. D {\bf 32}, 1871 (1985).
\item \label{kn:hu} E. Calzetta, S. Habib, and B. L. Hu, Phys. Rev. D {\bf 37},
2901 (1988).
\item \label{kn:kandrup0} H. E. Kandrup, Phys. Rev. D {\bf 37}, 2165 (1988).
\item \label{kn:graziani} F. R. Graziani, Phys. Rev. D {\bf 38}, 1122 (1988).
\item \label{kn:vlasov} A. A. Vlasov, {\em Statistical distribution functions}
(Nauka, Moscow, 1966) (in Russian).
\item \label{kn:birdav} N. D. Birrell and P. C. W. Davies, {\em Quantum Fields
in Curved Space} (Cambridge University Press, Cambridge, 1982).
\item \label{kn:fonarevthes} O. A. Fonarev, Ph.D. Thesis, Tomsk University,
Tomsk, USSR, 1991 (in Russian).
\item \label{kn:habib1} S. Habib and H. E. Kandrup, Ann. Phys. (N. Y.) {\bf
191}, 335 (1989).
\item \label{kn:stewart} J. M. Stewart, {\em Non--Equilibrium Relativistic
Kinetic Theory} (Lecture Notes in Physics, Vol. 10) (Springer, Berlin, 1971).
\item \label{kn:yano} K. Yano and Sh. Ishihara, {\em Tangent and Cotangent
Bundles: Differential Geometry} (Pure and Applied Mathematics: A Series of
Monographs and Textbooks) (Marcel Dekker, Inc., New York, 1973).
\item \label{kn:kobayashi} Sh. Kobayashi and K. Nomizu, {\em Foundations of
Differential Geometry}, Vol. 1 (Interscience Publishers, New York, 1963).
\item \label{kn:wigner} E. P. Wigner, Phys. Rev. {\bf 40}, 749 (1932).
\item \label{kn:degroot} S. R. de Groot, W. A. van Leeuwen, and Ch. G. van
Weert, {\em Relativistic Kinetic Theory} (North--Holland, Amsterdam, 1980).
\item \label{kn:fonarev1} O. A. Fonarev, Izv. VUZov. Fizika (USSR) {\bf 9}, 47
(1990).
\item \label{kn:carruthers} P. Carruthers and F. Zachariasen, Rev. Mod. Phys.
{\bf 55}, 245 (1983).
\item \label{kn:fock} V. A. Fock and D. Ivanenko, Z. Phys., {\bf 54}, 798
(1929).
\item \label{kn:bjorken} J. D. Bjorken and S. D. Drell, {\em Relativistic
Quantum Mechanics} (McGraw--Hill, 1964).
\item \label{kn:duffin} R. J. Duffin, Phys. Rev. {\bf 54}, 1144 (1938).
\item \label{kn:kemmer} N. Kemmer, Proc. Roy. Soc. (London) A {\bf 17}, 91
(1939).
\item \label{kn:ignat} A. V. Zakcharov and Yu. G. Ignat'ev, in {\em
Gravitatsiya i teriya otnositel'nosti (Gravitation and Relativity)} {\bf 13},
49 (1976) (Kazan University Press, Kazan, USSR) (in Russian).
\item \label{kn:heller} E. J. Heller, J. Chem. Phys. {\bf 65}, 1289 (1976).
\item \label{kn:habib2} S. Habib, Phys. Rev. D {\bf 42}, 2566 (1990).
\item \label{kn:dixon} W. G. Dixon, in {\em Proceedings of the International
School of Physics "Enrico Fermi"}, Varenna on Lake Como, Italy, 1976, edited by
J. Ehlers (North--Holland, 1979).
\item \label{kn:habib3} I am grateful to the referee for this comment.
\item \label{kn:winter1} J. Winter, J. Phys. (Paris) C {\bf 6}, 53 (1984).
\item \label{kn:israel} W. Israel, Lett. Nuovo Cim. {\bf 7}, 860 (1973).
\item \label{kn:heinz} U. Heinz, Phys. Rev. Lett. {\bf 51}, 351 (1983).
\item \label{kn:kandrup1} B. L. Hu and H. E. Kandrup, Phys. Rev. D {\bf 35},
1776 (1987).
\item \label{kn:kandrup2} H. E. Kandrup, Phys. Lett. B {\bf 202}, 207 (1988).
\item \label{kn:kandrup3} H. E. Kandrup, Phys. Rev. D {\bf 37}, 3505 (1988).
\item \label{kn:hu2} E. Calzetta and B. L. Hu, Phys. Rev. D {\bf 37}, 2878
(1988).
\item \label{kn:pirk} K. Pirk and G. B$\ddot{o}$rner, Class. Quantum Grav. {\bf
6}, 1855 (1989).
\item \label{kn:fonarev3} O. A. Fonarev, Phys. Lett. A {\bf 152}, 153 (1991).
\item \label{kn:fonarevunp} O. A. Fonarev (unpublished).
\item \label{kn:messer} J. Messer, Class. Quantum Grav. {\bf 4}, 1383 (1987).
\item \label{kn:serre} J.--P. Serre, {\em Lie algebras and Lie groups: Lectures
given at Garvard University} (Benjamin, New York, 1965).
\item \label{kn:akhiezer} A. I. Akhiezer and S. V. Peletminskii, {\em Methods
of Statistical Physics} (Pergamon Press, 1981).
\end{list}
\end{document}